\documentclass[12pt,letterpaper]{article}
\pdfoutput=1
\usepackage{jheppub}
\usepackage{amsmath}
\usepackage{bm}
\usepackage[utf8]{inputenc}

\hypersetup{
    colorlinks=true,   % false: boxed links; true: colored links
    linkcolor=red,     % color of internal links (change box color with linkbordercolor)
    citecolor=blue,    % color of links to bibliography
    filecolor=magenta, % color of file links
    urlcolor=blue      % color of external links
}

\toccontinuoustrue

\def\eq{\,{=}\,}
\def\Nc{N}
\def\half{\tfrac{1}{2}}
\def\rh{\rho_{\rm h}}
\def\rc{\rho_*}
\def\R{\mathbb R}
\def\Nfour{{\mathcal N}\,{=}\,4}
\def\O{{O}}

\def\ec{\epsilon_{\rm c}}
\def\Tc{T_{\rm c}}
\def\Tf{T_{\rm f}}
\def\tc{t_{\rm c}}
\def\CV{C_{\mathcal V}}
\def\Z{{\mathbb Z}}

\def\tr{\,{\rm tr}\,}

\title		{Large $\bm N$ phase transitions and the fate
		of small Schwarzschild-AdS black holes}
\author		{Laurence G.~Yaffe}
\emailAdd	{yaffe@phys.washington.edu}
\affiliation	{Department of Physics, University of Washington,
		Seattle, WA 98195-1560, USA}

\abstract
    {%
    Sufficiently small Schwarzschild-AdS black holes in asymptotically
    global AdS$_5 \times S^5$ spacetime are known to become dynamically
    unstable toward deformation of the internal $S^5$ geometry.
    The resulting evolution of such an unstable black hole is related,
    via holography, to the dynamics of supercooled plasma
    which has reached the limit of metastability
    in maximally supersymmetric large-$\Nc$ Yang-Mills theory
    on $\R \times S^3$.
    Puzzles related to the resulting dynamical evolution
    are discussed, with a key issue involving differences between
    the large $\Nc$ limit in the dual field theory and typical
    large volume thermodynamic limits.
    }

\begin{document}
\maketitle

\section	{Introduction}
\label		{sec:intro}

AdS/CFT duality relates the dynamics of maximally supersymmetric
SU($\Nc$) Yang-Mills theory ($\Nfour$ SYM), in the limit of
large $\Nc$ and large 't Hooft coupling $\lambda$, to
classical supergravity on asymptotically AdS$_5 \times S^5$
spacetimes \cite{Maldacena:1997re,Witten:1998qj,Gubser:1998bc}.
If the field theory is defined on a spatial three-sphere,
then the relevant dual geometries are asymptotic to global
AdS$_5$ (times $S^5$), whose conformal boundary may be taken
to be $\mathbb R \times S^3$.
One may understand the existence of a first order
confinement/deconfinement transition in $\Nfour$ SYM,
on $\mathbb R \times S^3$,
as a transition between two different gravitational solutions
both satisfying the required asymptotic behavior,
namely the vacuum geometry AdS$_5$ and the
Schwarzschild-AdS$_5$ black hole (both times $S^5$) \cite{Witten:1998zw}.
Sufficiently large Schwarzschild-AdS black holes provide the dual
description of thermal equilibrium states in $\Nfour$ SYM
above the deconfinement transition.

Smaller Schwarzschild-AdS black holes (BH)
should have a dual interpretation
involving non-equilibrium states in $\Nfour$ SYM\@.
In particular, it is known that Schwarzschild-AdS$_5 \times S^5$
black holes below a critical size become dynamically unstable
\cite{Hubeny:2002xn,Buchel:2015gxa,Dias:2015pda}.
The instability involves deformations of the internal $S^5$ geometry
and is thought to lead to localization of the black hole on the
internal space
\cite{Banks:1998dd,Peet:1998cr,Hubeny:2002xn,Dias:2016eto}.

The goal of this paper is to highlight a number of puzzling issues
in this presumed evolution scenario.
These include the inequivalence between canonical and
microcanonical descriptions of equilibrium states,
possible non-perturbative (finite $\Nc$) instabilities
in small but locally stable black holes,
and the connections (or lack thereof) between the dynamics of
unstable Schwarzschild-AdS black holes and spinodal decomposition
in typical first order phase transitions.
The possible validity of two alternative evolution scenarios is
also discussed.
One involves the potential existence of other stationary
supergravity solutions to which the dynamical evolution of
unstable Schwarzschild-AdS black holes might asymptote.
A more radical possibility is that this dynamical evolution
could fail to asymptote to any stationary solution.
This would indicate a failure of the corresponding non-equilibrium
initial states in $\Nfour$ SYM to thermalize.
Such lack of thermalization, in a strongly coupled field theory,
would be somewhat analogous to the phenomena of many body
localization present in certain condensed matter systems
\cite{PhysRevB.82.174411,
PhysRevB.83.094431,
doi:10.1146/annurev-conmatphys-031214-014726,
Chandran:2013xwa}.

A dissatisfying feature of the discussion in this paper is the lack
of crisp answers to some of the puzzles and speculative possibilities
considered.
Despite this, it is hoped that there is value in bringing attention
to the issues involved, on which future work may shed greater light.
In an effort to make the presentation largely self-contained,
section \ref{sec:N=4} summarizes key aspects of $\Nfour$ SYM
thermodynamics and properties of Schwarzschild-AdS black holes.
Section \ref{sec:1stord} reviews the behavior of typical
systems with a first order phase transition, focusing on
cooling dynamics in which a system initially in equilibrium
in the hot phase slowly loses energy, enters a metastable
supercooled state, undergoes spinodal decomposition upon
reaching the limit of metastability, and ultimately re-thermalizes.
This material is relevant background for the subsequent discussion.
Known results on the instability of small Schwarzschild-AdS black holes,
other deformed and localized black hole solutions 
\cite{Dias:2015pda,Dias:2016eto},
and the presumed dynamical scenario which could lead from an unstable 
Schwarzschild-AdS black hole to the known localized BH solutions
is summarized in section \ref{sec:smallBH}.
Section \ref{sec:scenarios}, discussing various issues in
this evolution scenario, as well as possible alternatives,
is the heart of this paper.
As noted above, open questions significantly outnumber clear answers.
The final section \ref{sec:conclusion} contains concluding remarks.

\section	{$\Nfour$ SYM on $\R \times S^3$ and AdS black holes}
\label		{sec:N=4}

We consider maximally supersymmetric $SU(\Nc)$ Yang-Mills theory
defined on $\R \times S^3$ with the spatial three-sphere having
radius $L$, and focus on the limit of large $\Nc$.
The finite spatial volume acts as an infrared cutoff and introduces
the length $L$ into the conformal field theory.
The curvature of the three-sphere induces a non-vanishing vacuum
Casimir energy, proportional to the number of degrees of freedom
and scaling as $1/L$,%
\footnote
    {%
    As emphasized in Ref.~\cite{Assel:2015nca},
    the Casimir energy on $\R \times S^3$
    is inherently ambiguous.
    Adding the counterterm
    $
	\int d^4x \, \sqrt g \, R^2
    $
    to the theory,
    with an arbitrary finite coefficient,
    shifts the vacuum energy on $\R \times S^3$.
    The value (\ref{eq:Casimir}) results from 
    calculations using $\zeta$-function regularization
    at zero coupling \cite{Birrell:1982ix}, as well as
    holographic renormalization
    \cite{Balasubramanian:1999re}
    in the gravitational dual.
    See also Refs.~\cite{Lorenzen:2014pna,Hirano:2017rbq}.
    }
\begin{equation}
    \lim_{\Nc\to\infty} \frac {E_{\rm vac}}{\Nc^2}
    \equiv
    \frac{3}{16L} \,.
\label{eq:Casimir}
\end{equation}

The spectrum of the theory is discrete,
and the density of states has a finite limit as $\Nc\to\infty$.
The Boltzmann sum representation of the canonical partition function
converges at sufficiently low temperatures, and the resulting
equilibrium state may be regarded as a thermal gas of glueballs
(gauge invariant excitations created, at large $\Nc$,
by single trace operators).
Since contributions to the pressure, or free energy,
due to a thermal gas of excitations scales as $\O(\Nc^0)$,
whereas the vacuum energy grows quadratically with $\Nc$,
the free energy in this ``confined'' phase is just
\begin{equation}
    F_{\rm conf}(T)
    = E_{\rm vac} + \O(\Nc^0) \,.
\label{eq:free energy}
\end{equation}
The entropy $S \equiv -\partial F/\partial T$ and
heat capacity $\CV \equiv \partial E/\partial T$
both scale as $\O(\Nc^0)$ in this phase,
and the $\Z_{\Nc}$ center symmetry of $\Nfour$ SYM is not spontaneously broken.

Above a transition temperature proportional to $1/L$,
\begin{equation}
    \Tc \equiv \frac {t_{\rm c}(\lambda)}{L} \,,
\end{equation}
the equilibrium state of the theory is a ``deconfined'' non-Abelian plasma
in which
the $\Z_{\Nc}$ center symmetry is spontaneously broken
(in the large $\Nc$ limit), 
and both the entropy and heat capacity are $\O(\Nc^2)$.
Quantitatively,
\begin{align}
    \tc(\lambda) =
    \begin{cases}
	1/\ln(7{+}4\sqrt 3) \approx 0.380 \,, & \lambda \to 0;
    \\
	3/(2\pi) \approx 0.477 \,, & \lambda \to \infty,
    \end{cases}
\end{align}
with the weak-coupling value determined by a perturbative reduction
to a solvable matrix model
\cite{Sundborg:1999ue,Aharony:2003sx}
and the strong-coupling value from the holographic description
reviewed below.
The transition between the confined and deconfined phases is a
genuine thermodynamic phase transition in the
$\Nc \to \infty$ limit, and is
first order with a non-vanishing and $\O(\Nc^2)$
latent heat (or discontinuity in the internal energy),
\begin{equation}
    \Delta E \equiv
    \lim_{T \to \Tc^+} E(T) -
    \lim_{T \to \Tc^-} E(T) \,,
\end{equation}
in both the weak coupling limit, $\lambda \equiv g^2 \Nc \to 0$,
and at strong coupling, $\lambda \gg 1$.%
\footnote
    {%
    Whether the deconfinement transition remains first order
    at small but non-zero coupling in $\Nfour$ SYM
    remains an open question
    \cite{Aharony:2003sx,Aharony:2005bq,Mussel:2009uw}.
    }

As discussed by Witten \cite{Witten:1998zw} and many other authors,
the above features of $\Nfour$ SYM thermodynamics,
in the strong coupling and large $\Nc$ limits,
have a simple dual gravitational description.
The geometry dual to the $\Nfour$ SYM vacuum state is
AdS$_5 \times S^5$.
A convenient form for the metric on this space is
\begin{equation}
    ds^2
    =
    -(1{+}\rho^2) \, dt^2
    + L^2 \left[
	\frac{d\rho^2}{1{+}\rho^2}
	+ \rho^2 \, d\Omega_3^2
	+ d\Omega_5^2 
    \right] ,
\label{eq:AdS}
\end{equation}
where the radial coordinate $\rho$ runs from 0 to $\infty$ at the AdS boundary.
In the gravitational description,
the length $L$ is both the AdS curvature scale and the
radius of the five-sphere.
The total energy,
extracted from the boundary stress tensor \cite{Balasubramanian:1999re},
is
\begin{equation}
    E_{\rm vac}
    = \frac {3\pi L^2}{32 \, G_5} 
    = \frac {3 \Nc^2}{16 L} \,,
\label{eq:Evac}
\end{equation}
where the last form uses the holographic relations
\begin{equation}
    \Nc^2
    = \frac {\pi L^3} {2\, G_5}
    = \frac {\pi^4 L^8} {2\, G_{10}}
    \,,
\label{eq:G}
\end{equation}
connecting the 5D and 10D Newton's constants $G_5$ and $G_{10}$ to
the rank of the $SU(\Nc)$ gauge group of the dual SYM theory.
Henceforth,
these relations will be used routinely to eliminate $G_5$ and $G_{10}$
and express results in terms of $\Nc$.
The Euclidean signature version of this geometry,
\begin{equation}
    ds^2
    =
    (1{+}\rho^2) \, dt^2
    + L^2 \left[
	\frac{d\rho^2}{1{+}\rho^2}
	+ \rho^2 \, d\Omega_3^2
	+ d\Omega_5^2 
    \right] ,
\end{equation}
with the periodic identification $t \eq t {+} \beta$,
provides the gravitational description of the thermal equilibrium state
at temperature $T \equiv \beta^{-1}$
in the confined phase of $\Nfour$ SYM theory.
This geometry is referred to as ``thermal AdS''.
The boundary stress-energy associated with this geometry gives the
leading $\O(\Nc^2)$ contributions to the field theory stress-energy tensor,
which is completely temperature independent
and equals the vacuum Casimir stress-energy.
On the gravitational side,
this temperature independence
reflects the fact that the periodic identification introducing
temperature has no effect whatsoever on local aspects of the geometry.
To see subleading $\O(\Nc^0)$ thermal contributions to the stress-energy,
or related thermodynamic observables,
one must include effects of quantum fluctuations in
the geometry.

The geometry (\ref{eq:AdS}) has no event horizon and hence vanishing
gravitational entropy.
Consequently, the free energy $F \equiv E {-} TS$ 
coincides with the vacuum energy,
\begin{equation}
    F_{\hbox{\scriptsize thermal-AdS}}
    %%% = \frac {3\pi L^2}{32 \, G} 
    = \frac {3 \Nc^2}{16 L} \,,
\label{eq:Fthermal}
\end{equation}
up to subleading $\O(\Nc^0)$ corrections.
This is in accord with the field theory entropy
(in the confined phase)
scaling as $\O(\Nc^0)$, and the free energy having the
form (\ref{eq:free energy}).

Equilibrium states in the deconfined phase of $\Nfour$ SYM have
a dual gravitational description in terms of
Schwarzschild black holes in asymptotically AdS$_5 \times S^5$
spacetimes \cite{Witten:1998zw}.
These ``AdS-BH'' geometries may be described by the metric
\begin{equation}
    ds^2
    =
    -f(\rho) \, dt^2
    + L^2 \left[
	\frac{d\rho^2}{f(\rho)}
	+ \rho^2 \, d\Omega_3^2
	+ d\Omega_5^2 
    \right] ,
\label{eq:AdSBH}
\end{equation}
where
\begin{equation}
    f(\rho) \equiv 1 + \rho^2 - (1+\rh^2) \> \frac{\rh^2}{\rho^2} \,.
\end{equation}
The dimensionless parameter $\rh$ controls the black hole size.
The black hole horizon is located at $\rho \eq \rh$
and the geometry reduces to that of global AdS$_5 \times S^5$
at $\rh \eq 0$.
The horizon area
$
    A = 2\pi^5 L^8 \rh^3
$,
and the associated black hole entropy is
\begin{equation}
    S \equiv \frac A{4G_{10}}
    %%% = \frac {\pi^2 L^3}{2G_5} \, \rh^3
    = \pi \Nc^2 \, \rh^3 \,.
\end{equation}
The horizon temperature
\begin{equation}
    T \equiv \frac {\kappa}{2\pi} 
    = \frac{\rh^{-1} + 2\rh}{2\pi L} \,,
\label{eq:T}
\end{equation}
where
$
    \kappa \equiv \half f'(\rh) / L
    %%% = (\rh^{-1} {+} 2\rh)/L
$
is the surface gravity at the horizon.
The temperature (\ref{eq:T}) has a minimum value
\begin{equation}
    T_{\rm min} = \frac {\sqrt 2}{\pi L}
\end{equation}
at $\rh \eq 1/\sqrt 2$, and diverges in the limit of both large
and small $\rh$.
Inverting the relation (\ref{eq:T}) gives the black hole size
as a double-valued function of temperature,
\begin{equation}
    \rh = \half \left[ \pi LT \pm \sqrt{(\pi LT)^2 - 2 } \right] .
\end{equation}
The $+$ branch is referred to as describing
``large'' AdS black holes,
while the $-$ branch describes ``small'' AdS black holes.

\begin{figure}
\begin{center}
\includegraphics[scale=0.59]{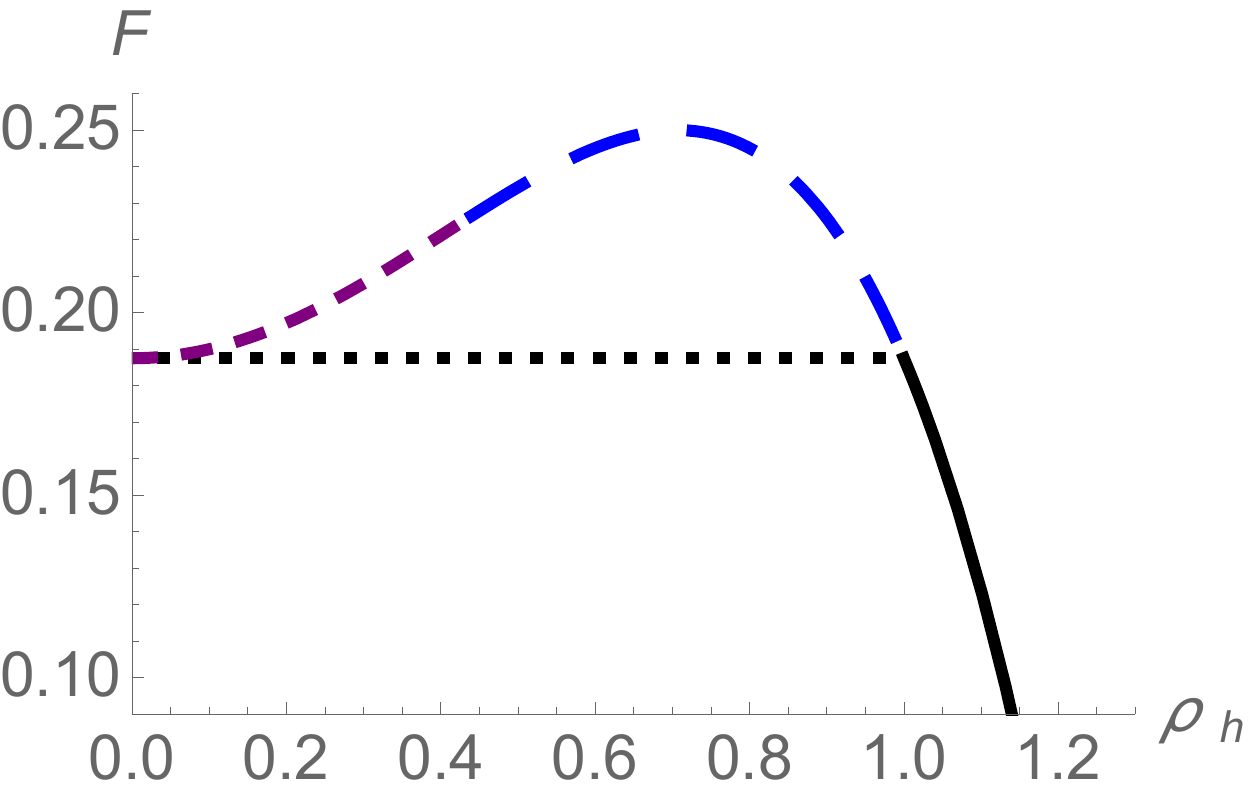}
\hfill
\includegraphics[scale=0.59]{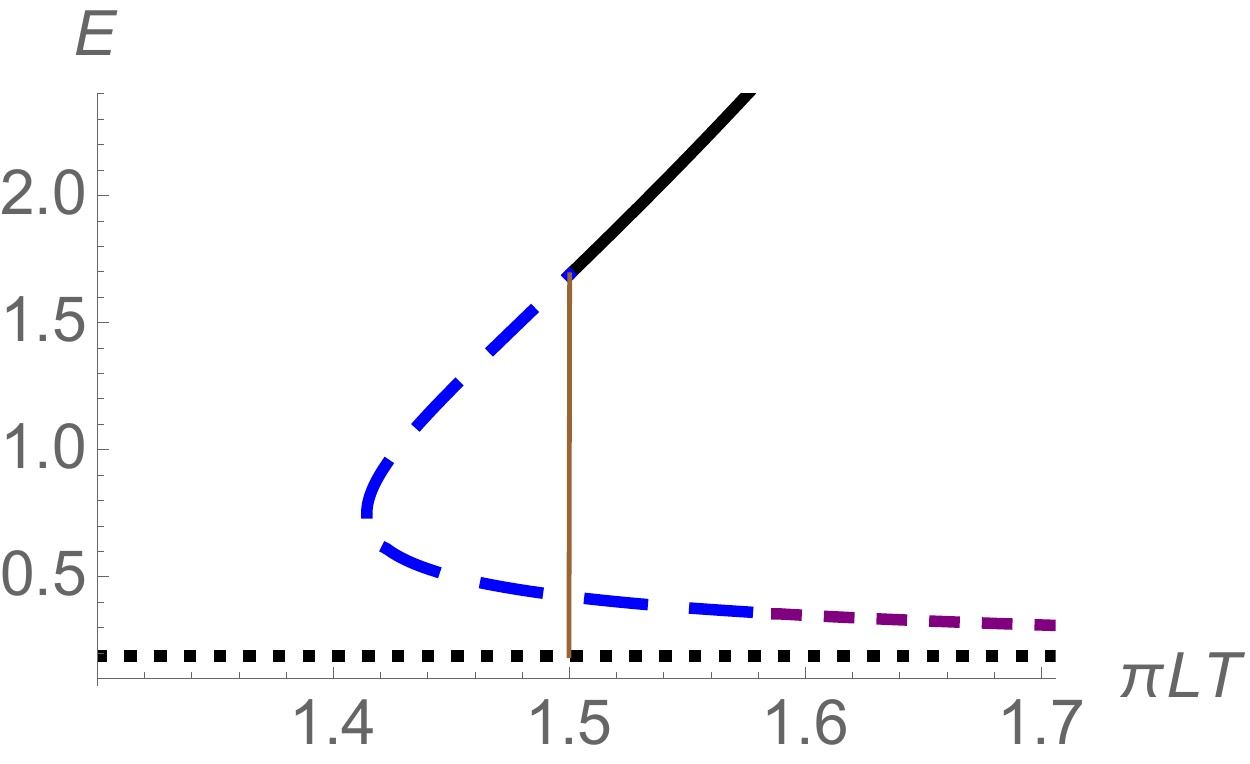}
\end{center}
\vspace*{-1em}
\caption
    {%
    Left panel:
    Free energy of asymptotically AdS$_5 \times S^5$
    Schwarzschild black holes,
    in units of $\Nc^2/L$,
    as a function of the black hole size $\rh$.
    The solid black portion of the curve with $\rh > 1$ represents
    stable equilibrium states of deconfined $\Nfour$ SYM plasma.
    The long dashed blue portion of the curve with $\rc < \rh < 1$
    represents locally stable states of supercooled plasma, while
    the short dashed purple portion with $\rh < \rc$ corresponds to
    locally unstable states.
    The dotted horizontal line shows the free energy of thermal-AdS\@.
    Right panel:
    Energy of AdS-Schwarzschild black holes, and thermal-AdS,
    in units of $\Nc^2/L$ and plotted as a function of temperature.
    The thin vertical line marks the transition temperature.
    Different markings on the curves have the same meanings
    as in the left panel.
    \label{fig:F}
    }
\end{figure}

The energy of the Schwarzschild-AdS solution (\ref{eq:AdSBH}),
extracted from the boundary stress-energy tensor
\cite{Balasubramanian:1999re}, is
\begin{equation}
    E
    %%% = \frac {3\pi L^2}{32 G} \, (1 + 2 \rh^2)^2
    = \frac {3 \Nc^2}{16 L} \, (1 + 2 \rh^2)^2 \,,
\label{eq:E}
\end{equation}
and the corresponding free energy
\begin{equation}
    F \equiv E - T \, S
    %%% = \frac{\pi L^2}{8 G} \, \big[ \tfrac 34 + \rh^2(1-\rh^2) \big]
    = \frac{\Nc^2}{4L} \, \big[ \tfrac 34 + \rh^2(1-\rh^2) \big] \,.
\label{eq:F}
\end{equation}
The left panel of Figure \ref{fig:F} shows a plot of the
free energy (\ref{eq:F}) together with the constant value (\ref{eq:Fthermal})
of (the $\O(\Nc^2)$ part of) the thermal AdS free energy.
The right panel shows the internal energy (\ref{eq:E}) of the AdS black hole,
along with the constant value of the thermal AdS energy (\ref{eq:Evac}),
as a function of temperature.

The pressure $p$
coincides with
$
    E/(3\mathcal V) = E/(6\pi^2 L^3)
$,
as required for a conformal theory with a traceless stress-energy tensor
(on $\R \times S^3$).
This agrees, as it must, with the thermodynamic definitions
$
    p
    = \left. -\frac {dE}{d\mathcal V} \right|_{S}
    = \left. -\frac {dF}{d\mathcal V} \right|_{T}
$.
For $\rh \gg 1$, the pressure approaches that of strongly coupled
$\Nfour$ SYM plasma in flat space,
$
    p \sim
    \frac{\Nc^2}{8\pi^2} \, \rh^4 \, L^{-4} =
    \frac{\pi^2}{8} \, \Nc^2 \, T^4
$.

In a canonical description of thermodynamics,
genuine equilibrium states are global minima of the free energy.
The black hole free energy (\ref{eq:F}) falls below the
thermal AdS free energy (\ref{eq:Fthermal}) only when $\rh > 1$.
The point $\rh \eq 1$ lies on the large BH branch and corresponds
to a transition temperature 
\begin{equation}
    \Tc = \frac 3{2\pi L} \,,
\end{equation}
where the nature of the equilibrium state switches between confined
and deconfined phases.
The transition is first order with two distinct equilibrium states,
thermal AdS (confined) and the AdS-BH (deconfined),
co-existing at $\Tc$.
The internal energy is discontinuous, with
\begin{equation}
    E_{\rm c}^- \equiv \lim_{T\to\Tc^-} E(T) = \frac{3\Nc^2}{16L} \,,\qquad
    E_{\rm c}^+ \equiv \lim_{T\to\Tc^+} E(T) = \frac{27\Nc^2}{16L} \,,
\end{equation}
and latent heat
$
    \Delta E = \frac {3 \Nc^2}{2 L}
$.
Examination of the expectation value of the (dual description of the)
Polyakov loop confirms that the $\Z_{\Nc}$ center
symmetry is unbroken in the thermal AdS geometry, but is
spontaneously broken in the AdS-BH geometry \cite{Witten:1998zw},
in complete accord with the interpretation of the transition between
these geometries as a confinement/deconfinement phase transition.%
\footnote
    {%
    In this finite volume theory,
    the Polyakov loop expectation value
    $\langle \frac 1{\Nc} \tr \Omega \rangle$,
    in the large $\Nc$ limit,
    may be defined via the large $\Nc$ factorization relation,
    $
	\lim_{\Nc \to \infty}
	\langle |\frac 1{\Nc} \tr \Omega |^2 \rangle
	=
	|\langle \frac 1{\Nc} \tr \Omega \rangle |^2
    $.
    For equivalent alternative definitions and related discussion 
    see, for example, Refs.~\cite{Witten:1998zw,Aharony:2003sx}.
    }
Finally, AdS black holes on the large BH branch are dynamically stable;
all quasinormal mode frequencies lie in the lower half plane,
corresponding to exponentially damped behavior.
To summarize, the interpretation of large AdS-Schwarzschild black holes
with $\rh \ge 1$ as the holographic duals of equilibrium states
of deconfined non-Abelian $\Nfour$ SYM plasma
(inside a three-sphere,
in the large $\Nc$ and strong coupling limits) is well established.

For $\rh < 1$, the AdS-BH is no longer the minimum of the free energy
\cite{Hawking:1982dh}.
The heat capacity, given by
\begin{equation}
    \CV \equiv \frac{\partial E}{\partial T}
    = 3\pi \Nc^2 \, \rh^3 \; \frac{2\rh^2 + 1}{2\rh^2 -1} \,,
\label{eq:CV}
\end{equation}
diverges to $+\infty$ as $\rh \to 1/\sqrt 2$ from above,
corresponding to $T \to T_{\rm min}$,
and becomes negative on the small-BH branch with $\rh < 1/\sqrt 2$.
As the black hole size decreases below $\rh = 1/\sqrt 2$,
the global AdS black hole ceases to be locally dynamically stable when
$\rh$ reaches a critical value
\cite{Hubeny:2002xn,Buchel:2015gxa,Dias:2015pda},
\begin{equation}
    \rc \approx 0.4402373 \,.
\end{equation}
At $\rh \eq \rc$, one quasinormal mode frequency crosses
the real axis and, for $\rh < \rc$, moves into the upper half plane,
indicating an exponentially growing instability.
The unstable mode involves a deformation of the geometry involving
$\ell \eq 1$ harmonics on the internal $S^5$.
Further instabilities, involving higher harmonics on the $S^5$,
appear at a series of progressively smaller values of $\rh$
\cite{Hubeny:2002xn}.
At each instability threshold there is a zero mode in the
static fluctuation spectrum, signaling a bifurcation in the
space of static solutions.
The new branches of static solutions which appear
at the first two such bifurcations were studied numerically in
Ref.~\cite{Dias:2015pda}.
These new solutions have deformations which break the $SO(6)$ symmetry
of the internal $S^5$ down to $SO(5)$,
and have been termed ``lumpy'' AdS black holes.

Schwarzschild-AdS black hole geometries with $\rc < \rh < 1$
must be the gravitational dual of some states in the dual field theory.
The non-minimal free energy indicates that this geometry does not
represent a true equilibrium state in $\Nfour$ SYM\@.
Nevertheless,
the geometry is the smooth continuation of the thermodynamically
stable black hole to lower values of energy and,
as noted above, it remains locally dynamically stable.
It should be clear that this geometry provides the dual description
of deconfined $\Nfour$ SYM plasma which is supercooled below the
confinement/deconfinement transition.
The point $\rh \eq \rc$ represents the limit of local metastability
of the supercooled system.%
\footnote
    {%
    Supercooled phases,
    before they reach the limit of metastability,
    typically have non-perturbative instabilities involving
    nucleation and growth \cite{Langer:1969bc}.
    In $SU(\Nc)$ gauge theories, however,
    such nucleation probabilities
    vanish exponentially as $\Nc \to \infty$
    and may largely be ignored when considering the dynamics
    of SYM plasma at large $\Nc$.
    Section \ref{sec:scenarios} contains further discussion
    of this issue.
    \label{fn:bubbles}
    }

\section	{Cooling through first order phase transitions}
\label		{sec:1stord}

Before examining possible scenarios for the dynamical evolution of 
small AdS black holes which reach the limit of metastability,
we first review cooling dynamics at typical first order
phase transitions.
This will be useful background for the subsequent discussion.

\begin{figure}
\begin{center}
\includegraphics[scale=0.6]{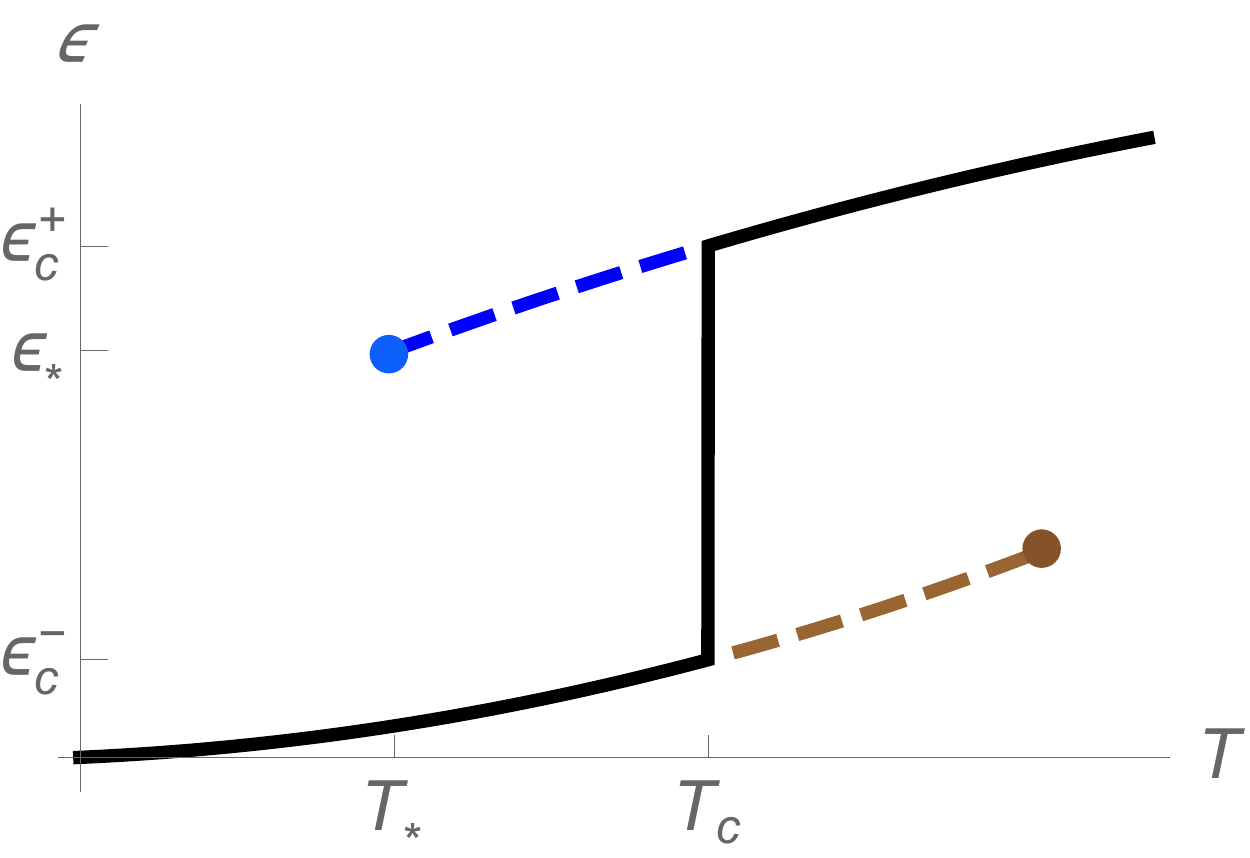}
\end{center}
\vspace*{-1em}
\caption
    {%
    Schematic form energy density $\epsilon$ as a function of
    temperature $T$ in a typical system with a first order phase
    transition at some temperature $\Tc$.
    The energy density jumps discontinuously from $\ec^-$ to $\ec^+$.
    Dashed curves show metastable supercooled states
    with energy densities and temperatures varying from
    $(\ec^+,\Tc)$ down to the limit of
    metastability at $(\epsilon_*,T_*)$, and corresponding
    metastable superheated states extending upward from
    $(\ec^-,\Tc)$.
    \label{fig:1stord}
    }
\end{figure}

Consider a system with a first order phase transition in the usual
thermodynamic limit of spatial volume $\cal V$ tending to infinity.
Assume, for simplicity, that this is not a symmetry-breaking
phase transition but instead something like a gas-to-liquid
transition for which there is a unique equilibrium state
on either side of the transition.
Figure \ref{fig:1stord} sketches the typical behavior of the 
energy density $\epsilon$ as a function of temperature,
showing a monotonically increasing function
with a discontinuous jump at the transition temperature $\Tc$.
Let $\ec^\pm \equiv \lim_{T \to \Tc^\pm} \epsilon(T)$
denote the energy densities at the transition,
with limits taken from the indicated side;
$\Delta\epsilon \equiv \ec^+ - \ec^-$ is the latent heat per unit volume.
The high temperature phase has a metastable supercooled continuation
below $\Tc$,
and likewise 
the low temperature phase has a metastable superheated
continuation above $\Tc$,
both indicated by dashed lines in the figure.
The limit of metastability of the supercooled system
lies at some temperature $T_*$ and energy density $\epsilon_*$.

Given the assumed absence of symmetry breaking,
at temperatures other than $\Tc$ there is a unique equilibrium state.
Uniqueness of the equilibrium state implies,
for example, that local (or compactly supported) observables have
no sensitivity to the choice of boundary conditions placed on
the spatial boundary of the theory before the volume
$\cal V$ is sent to infinity.

Precisely at $\Tc$ there are multiple equilibrium states.
These include the two ``pure phases'' with energy densities
$\ec^\pm$ which are the limits of the unique
equilibrium states on either side of the transition.
Let $\rho_\pm$ denote the statistical density matrices
of these equilibrium states.
Possible equilibrium states
also include statistical mixtures of the
two pure phases,%
\footnote
    {%
    In infinite volume, an ``equilibrium state'' means
    a probability measure which satisfies the Dobrushin-Lanford-Ruelle
    conditions (see, e.g., Ref.~\cite{Bratteli:1996xq}),
    namely that when restricted to any finite subvolume with degrees
    of freedom outside the subvolume fixed, the measure reduces to
    the canonical Gibbs measure conditioned on the fixed degrees of freedom.
    }
\begin{equation}
    \rho = x \, \rho_+ + (1{-}x) \, \rho_- \,,
\label{eq:mixed}
\end{equation}
with $x \in (0,1)$.
Such ``mixed states'' form the interior of a convex domain whose
extremal points are the pure phases.%
\footnote
    {%
    More generally
    if some discrete symmetry, say $\mathbb Z_p$, is unbroken on
    one side of the transition and spontaneously broken
    on the other side, then there can be $p{+}1$
    pure phases at $\Tc$, with the most general mixed phase
    an arbitrary convex linear combination of these 
    $p{+}1$ pure phases.
    }
A key point is that equilibrium states at $\Tc$ exist
with any desired energy density
in between the extremal values,
$\epsilon \in [\ec^-,\ec^+]$.

Cluster decomposition provides a diagnostic indicating whether a
given equilibrium state is pure or mixed.
For later purposes,
note that the usual statement of cluster decomposition,
\begin{equation}
    \lim_{|x{-}y|\to\infty}
    \left[
    \left\langle \mathcal O(x) \mathcal O(y) \right\rangle
    -
    \left\langle \mathcal O(x) \right\rangle
    \left\langle \mathcal O(y) \right\rangle
    \right]
    =
    0
\label{eq:cluster1}
\end{equation}
for some local observable $\mathcal O(x)$ (in infinite volume),
implies the integrated form,
\begin{equation}
    \lim_{|\mathcal R| \to \infty}
    \left[
    \left\langle \left(
	\textstyle \frac 1{|\mathcal R|} \int_{\mathcal R} \> \mathcal O(x)
    \right)^2 \right\rangle
    -
    \left\langle
	\textstyle \frac 1{|\mathcal R|} \int_{\mathcal R} \> \mathcal O(x)
    \right\rangle^2
    \right]
    =
    0 \,,
\label{eq:cluster2}
\end{equation}
where the region $\mathcal R$ is, for example, a ball of volume
$|\mathcal R|$.
In other words, cluster decomposition implies 
vanishing variance of intensive observables spatially averaged 
over increasingly large volumes in the thermodynamic limit.
Equilibrium states which satisfy cluster decomposition are
called ``extremal'' or ``pure'' states,%
\footnote
    {%
    This use of ``pure states'' in statistical physics,
    as a synonym of ``extremal'', is distinct
    from pure states in quantum mechanics.
    }
while equilibrium states which violate cluster decomposition are
mixed states of the form (\ref{eq:mixed}),
decomposable into a positively weighted average of extremal states.

One might expect further distinct ``phase-separated''
equilibrium states to also exist at $\Tc$, in which an interface
(or domain wall) is present on some planar surface, with properties
on one side approaching those of the pure phase $\rho_+$ while
properties on the other side approach those of $\rho_-$.
This is the case in four or more spatial dimensions,
where bounded transverse fluctuations of an interface allow
non-translationally invariant extremal equilibrium states
(satisfying cluster decomposition) to exist.%
\footnote
    {%
    For lattice theories in three space dimensions,
    non-translationally invariant equilibrium states exist below
    the interface roughening temperature
    \cite{PhysRevLett.38.993,PhysRevB.14.4978,2015JSP...159..958C}.
    The lower limit of four spatial dimensions applies to
    continuum theories with continuous rotation and translation
    invariance, which the following discussion assumes.
    }
Such states may be viewed as the limit of finite volume
equilibrium states in which one fixes degrees of freedom
on the boundary of the volume in a non-uniform
manner which pins the interface location on the boundary
and selects the desired volume fraction $x$
\cite{2015JSP...159..958C}.
But in three or fewer spatial dimensions, which the
following discussion assumes,
the transverse fluctuations in interface position
diverge as $\mathcal V \to \infty$.
As a result, it becomes completely indeterminate whether
some given spatial region $\mathcal R$,
of finite but arbitrarily large extent,
lies on one side of the interface or the other.
The probability that the interface runs through
any given region $\mathcal R$ vanishes as $\mathcal V \to \infty$.
Consequently, as probed by any local (or compactly supported)
observable, a phase-separated equilibrium state,
in three or fewer space dimensions,
is a mixed phase of the form (\ref{eq:mixed}),
and does not satisfy cluster decomposition.%
\footnote
    {%
    Note, however, that adding a non-translationally invariant
    perturbation to the system,
    such as a tiny gravity gradient in a fluid system, 
    can serve to localize an interface separating pure phases,
    thereby producing a non-translationally invariant phase-separated
    equilibrium state which does satisfy cluster decomposition.
    }

Dynamically, in the infinite volume limit,
an interface surface will continually fluctuate,
undergoing unbounded stochastic motion,
and never settle down to a well-defined position.
Interpreting the mixed ensemble (\ref{eq:mixed}) as a
phase-separated equilibrium state with a completely uncertain
interface location makes clear that these
non-extremal equilibrium states are also present in
a microcanonical description of thermodynamics.
Fixing the total energy, or rather energy density,
to lie in between the extremal values,
$\epsilon_{\rm c}^- < \epsilon < \epsilon_{\rm c}^+$,
determines the volume fraction $x$ but leaves the location
of the separating interface completely undetermined.

Given this understanding of possible equilibrium states,
consider a system which is initially prepared and fully
equilibrated at some temperature $T > \Tc$,
and then slowly cooled
(for example, by adiabatic expansion in a fluid system).
The cooling removes energy from the system and lowers
its temperature through $\Tc$, causing the system
to enter the metastable supercooled regime.
Within this regime,
assume that the cooling rate is small compared to
microscopic relaxation rates but is large compared to the
rate for bubble nucleation
(within some finite region of interest),
so that the system remains in the supercooled state
up to the limit of metastability,
whereupon it becomes unstable.
What happens next?

\begin{figure}
\begin{center}
\includegraphics[scale=0.6]{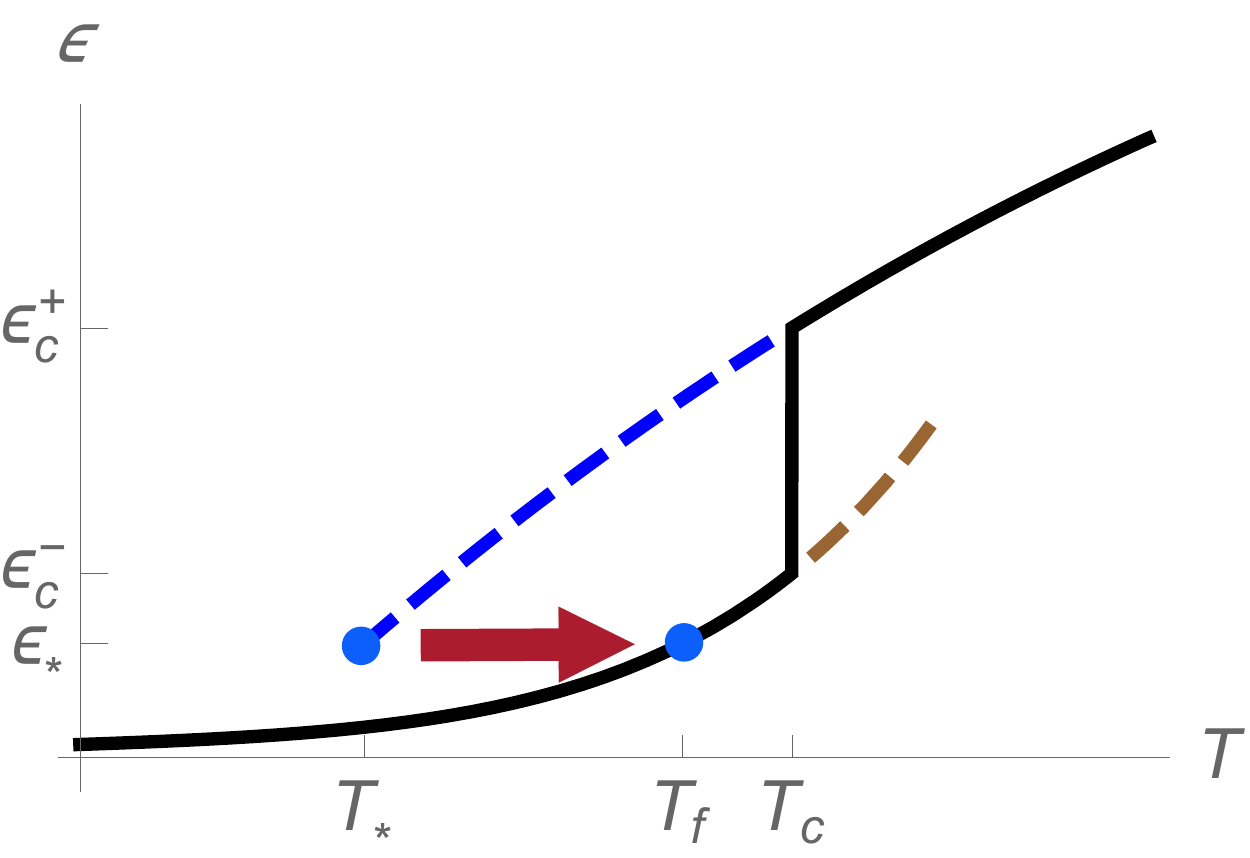}
\hfill
\includegraphics[scale=0.6]{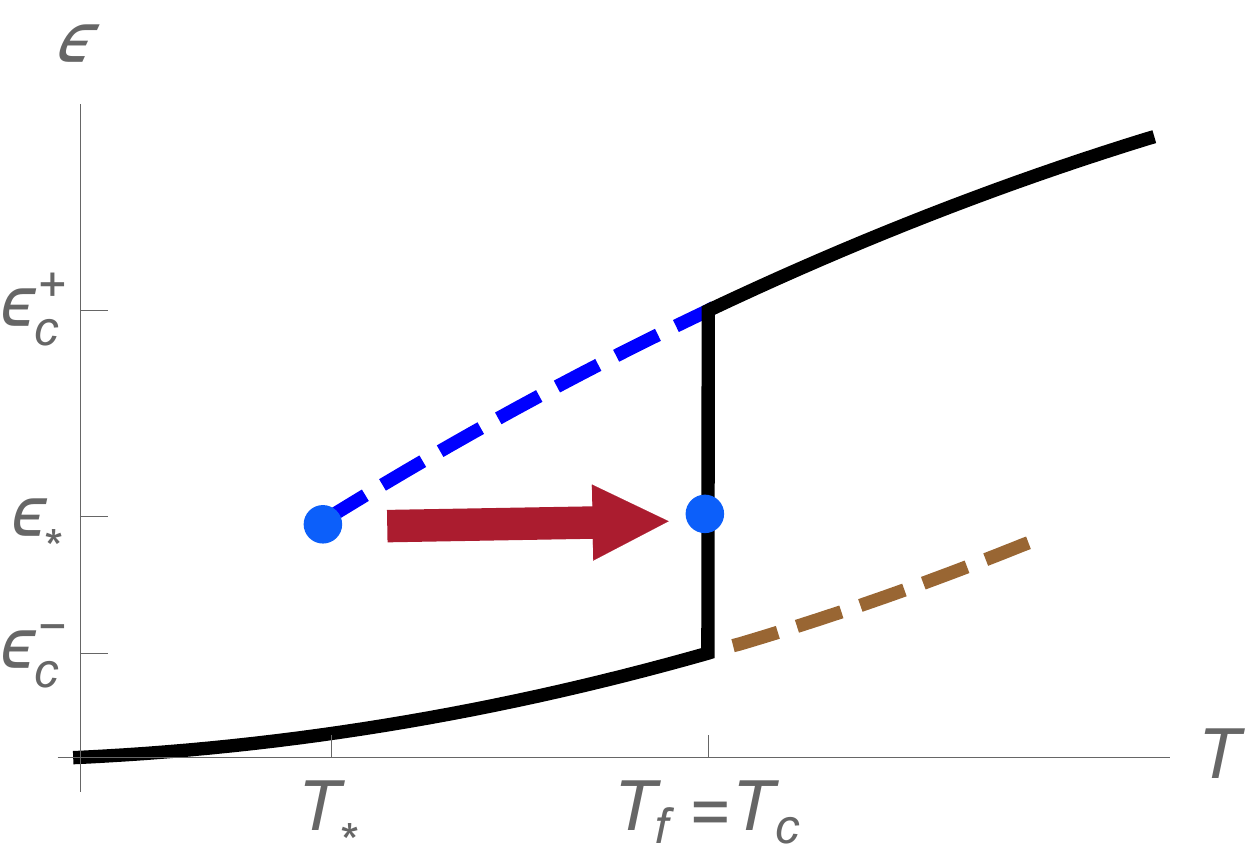}
\end{center}
\vspace*{-1em}
\caption
    {%
    Equilibration after spinodal decomposition.
    Left panel: $\epsilon_* < \ec^-$, leading to a final temperature
    $\Tf < \Tc$.
    Right panel: $\epsilon_* > \ec^-$, leading to a 
    mixed phase equilibrium state with $\Tf \eq \Tc$.
    \label{fig:rethermalize}
    }
\end{figure}

The subsequent evolution is termed \emph{spinodal decomposition}.
Typically, perturbations with a range of wavenumbers become
unstable in the homogeneous supercooled state
when $T < T_*$,
leading to growth of structure with intricate spatial patterns.
In, for example, solids with first order compositional phase
transitions, if a material undergoing spinodal decomposition is
cooled sufficiently rapidly then finely dispersed spatial microstructure
can become frozen in place, significantly changing material properties
in technologically useful fashions.
But suppose, instead, that the external cooling ceases at the onset
of spinodal decomposition, so that the system subsequently evolves
as an isolated system.
Although the details of the dynamics may be highly complex,
in any normal thermodynamic system the endpoint of the evolution
is easy to describe: the system re-thermalizes.
The energy density of the system is $\epsilon_*$
at the onset of spinodal decomposition and therefore,
since energy is conserved, the system must end up in an
equilibrium state with the same energy density $\epsilon_*$.

There are two characteristic possibilities, illustrated in
Fig.~\ref{fig:rethermalize}.
If $\epsilon_* < \ec^-$, then the system can equilibrate
to an equilibrium state on the low temperature side of the
phase transition, with a final temperature $\Tf$ which is below $\Tc$.
Alternatively, if $\epsilon_* > \ec^-$, then the system
must equilibrate to a mixed phase with 
energy density $\epsilon_*$ and final temperature $\Tf \eq \Tc$.
Given that $\epsilon_* < \ec^+$, it is impossible
for the system to equilibrate with a final temperature $\Tf > \Tc$.
The basic point to be emphasized is that,
if $\epsilon_* > \ec^-$,
then equilibration after spinodal decomposition \emph{must}
lead to a final state at temperature $\Tc$.

\section	{Supercooled $\Nfour$ plasma and small black holes}
\label		{sec:smallBH}

We now return to consideration of supercooled $\Nfour$ SYM plasma,
or small AdS black holes.
The basic scenario is the same as discussed above:
$\Nfour$ plasma is confined to a three-sphere of radius $L$
and initially in equilibrium at some temperature $T > \Tc$.
The three-sphere is slowly expanded.
As the radius $L$ increases,
the positive pressure of the plasma implies that the plasma does work and
loses energy.
When the energy diminishes below
$
    E_{\rm c}^+ \equiv {27 \Nc^2}/({16L})
$
the system passes into the supercooled regime.
As noted earlier (footnote \ref{fn:bubbles}),
homogeneous nucleation within the supercooled regime is
exponentially suppressed at large $\Nc$ and may be ignored.
The expansion continues until the plasma crosses the limit
of metastability at an energy of
\begin{equation}
    E_* \equiv \frac {3\Nc^2}{16L} \, (1+2\rho_{\rm c}^2)^2
    \approx
    \frac{0.361028 \, \Nc^2}L \,,
\label{eq:E*}
\end{equation}
and then ceases.
The peculiar form (\ref{eq:CV}) of the heat capacity shows that
the temperature is non-monotonic during the expansion,
initially decreasing and passing through $\Tc = 1.5/(\pi L)$,
reaching $T_{\rm min} \approx 1.414/(\pi L)$ and then
increasing to $T_* \approx 1.576/(\pi L)$.
Nothing dynamically significant happens at $T_{\rm min}$;
the pressure remains positive and energy is continually extracted
as the volume expands.
A local instability of the supercooled plasma arises when the
energy drops below $E_*$ (\ref{eq:E*}),
and the question we wish to consider is what
can be said about the endpoint of the subsequent evolution.
This is the same as asking what, in the dual gravitational
description, is the fate of a small Schwarzschild-AdS black hole
after it becomes unstable at $\rho < \rho_{\rm c}$?

This question, and closely related issues, have been considered previously.
The conventional expectation is that the horizon of the AdS black hole,
with $S^3 \times S^5$ topology,
becomes increasingly distorted on the $S^5$ in a manner
analogous to the Gregory-Laflamme instability of black strings
in asymptotically flat space \cite{Gregory:1993vy}.
A cascade of instabilities on different scales leads to the
development of fractal structure 
\cite{Lehner:2010pn,Lehner:2011wc}.
The classical gravity description breaks down when the length of
minimal cycles around the horizon reach the string scale $\ell_s$.
At this point, topology changing transitions which break thin
``necks'' in the horizon become possible.
The presumption has been that this process will lead to
one or more black holes with $S^8$ horizon topology, which
eventually merge and settle down to form a single stationary black hole,
localized on the $S^5$,
with a geometry which is invariant under the $SO(4)$ symmetry of
the spatial $S^3$ and at most an $SO(5)$ subgroup of the 
$SO(6)$ symmetry of the asymptotic $S^5$.

\begin{figure}
\begin{center}
\hspace*{10pt}
\includegraphics[trim=0 2pt 0 0,scale=0.447]{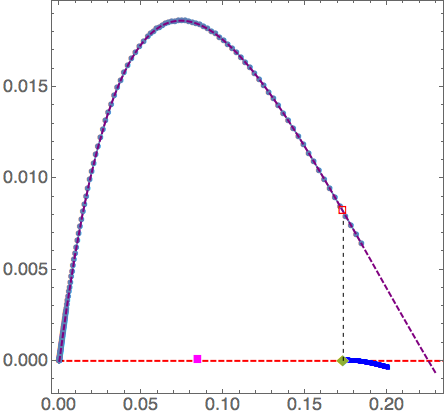}
\begin{picture}(0,0)(0,0)
    \put(-69,67.0){\footnotesize $\delta S/N^2$}
    \put(8,67.0){\footnotesize $\delta F \, L/N^2$}
    \put(-38,-4){\footnotesize $\delta E \, L/N^2$}
    \put(44,-4){\footnotesize $T \, L$}
\end{picture}
\hfill
\includegraphics[scale=1.00]{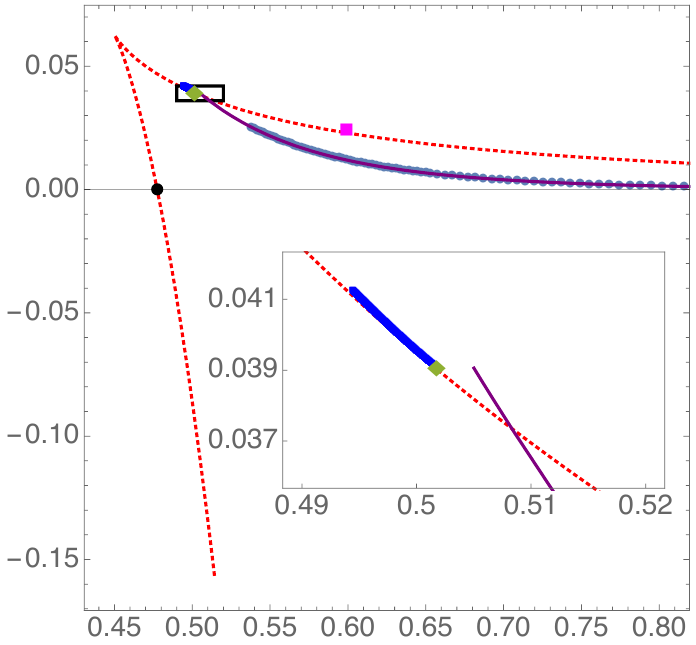}
\vspace*{10pt}
\caption
    {%
    Left: Entropy difference
    $\delta S \equiv S - S_{\textrm{Schw-AdS}}$
    of localized black holes relative to the
    corresponding Schwarzschild-AdS$_5 \times S^5$ black hole,
    as a function of energy excess
    $\delta E \equiv E - E_{\rm vac}$.
    Individual black data points show the additional entropy
    $\delta S$ of localized $S^8$ horizon topology black holes;
    the dashed purple curve through these points is a fit to the data.
    The horizontal dashed red line at $\delta S \eq 0$ represents
    Schwarzschild-AdS$_5 \times S^5$ black holes.
    The green diamond and pink square mark the $\ell \eq 1$ 
    and $\ell \eq 2$ bifurcations, respectively, of the
    AdS$_5 \times S^5$ black hole.
    The thick blue curve emerging from the $\ell \eq 1$ bifurcation
    shows lumpy $S^3 \times S^5$ horizon topology black holes.
    The thin dashed vertical line identifies the localized
    black hole with the same energy as the Schwarzschild-AdS
    black hole at the onset of instability.
    Right: Free energy excess $\delta F \equiv F - E_{\rm vac}$
    as a function of temperature.
    Different curves and symbols have the same meanings as in
    the left plot.
    The upper and lower red dotted curves show the free energy excess
    of small and large Schwarzschild-AdS black holes, respectively.
    The single black dot on the lower branch marks the phase
    transition point where the black hole free energy falls
    below that of thermal-AdS\@.
    The inset plot magnifies the region near the $\ell\eq 1$
    bifurcation.
    Localized $S^8$ black holes (black data points with
    purple fit to the data) have lower free energy than small
    Schwarzschild-AdS black holes for $T \gtrsim 0.585/L$.
    Plots courtesy of O.~Dias and reproduced from
    Ref.~\cite{Dias:2016eto}.
    \label{fig:localized}
    }
\end{center}
\end{figure}
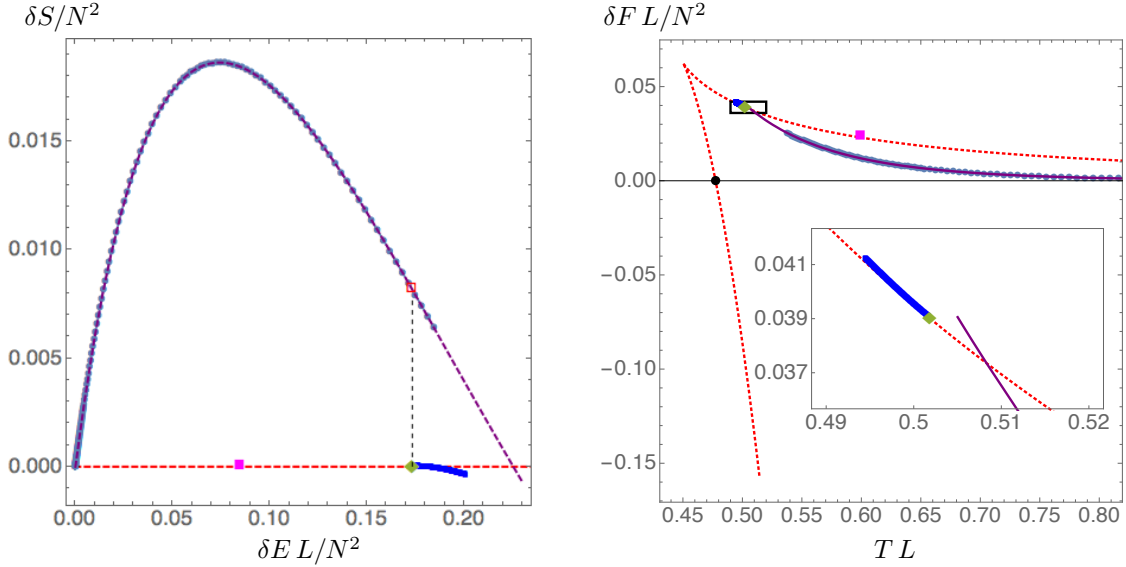

Much of this picture is implicit in early 1998 discussions
of AdS black holes and associated thermodynamics in a microcanonical
perspective by Banks, Douglas, Horowitz, and Martinec
\cite{Banks:1998dd}
and by Peet and Ross \cite{Peet:1998cr}.
The first explicit calculation of instability thresholds of
Schwarzschild-AdS$_5 \times S^5$ black holes was performed by
Hubeny and Rangamani \cite{Hubeny:2002xn}, who
emphasized that from a gauge theory perspective
these instabilities should be interpreted as indicating the existence
of phase transitions in the microcanonical ensemble at which the
$SO(6)_R$ symmetry of $\Nfour$ SYM is spontaneously broken.
As mentioned earlier, ``lumpy'' static black hole branches which
emerge from the first two bifurcations were constructed numerically by
Dias, Santos, and Way \cite{Dias:2015pda}.
For the leading ($\ell \eq 1$) bifurcation they 
found that the entropy decreases as the deformation increases,
so these deformed black holes are even less thermodynamically relevant
than undeformed Schwarzschild-AdS black holes at the same energy.
More recently \cite{Dias:2016eto}, these authors succeeded
in finding ``localized'' black hole solutions with $S^8$ horizon
topology and $SO(4) \times SO(5)$ symmetry.
The localized black holes have
higher entropy than Schwarzschild-AdS$_5 \times S^5$
black holes for energies below \cite{Dias:2016eto}
\begin{equation}
    E_R \approx 0.413 \, \Nc^2/L \,.
\label{eq:E_R}
\end{equation}
The localized BH with energy $E_*$ has a temperature
\begin{equation}
    T^{\rm loc}_* \approx 1.7098 / (\pi L) \,,
\end{equation}
higher than both $\Tc$ and $T_*$.
These localized solutions may eventually meet the branch of lumpy black
holes emerging from the $\ell \eq 1$ bifurcation at a cusp-like
topological transition point lying at yet higher energy.

The left panel of Figure~\ref{fig:localized}
shows the entropy difference
$\delta S \equiv S - S_{\textrm {Schw-AdS}}$
of these localized black holes
relative to that of an undeformed Schwarzschild-AdS black hole,
as a function of energy excess $\delta E \equiv E - E_{\rm vac}$.
The black data points with purple dashed fit to the data
represent localized black holes.
The horizontal red line at $\delta S = 0$ corresponds
to Schwarzschild-AdS black holes.
On this line, the green diamond at
$
    \delta E_* \equiv E_* - E_{\rm vac} \approx 0.1735 \, \Nc^2/L
$
indicates the $\ell = 1$ bifurcation (i.e., the metastability limit),
while the pink square shows the next $\ell = 2$ bifurcation.
The blue curve emerging from the $\ell = 1$ bifurcation represents
lumpy black hole solutions.
The intersection of the localized BH curve with the
horizontal red line at
$\delta E_R = E_R - E_{\rm vac} \approx 0.225 \, \Nc^2/L$
represents the point where the (extrapolated)
entropy of the localized solutions falls below
that of a Schwarzschild-AdS black hole.
(Plots courtesy of O.~Dias and reproduced from
Ref.~\cite{Dias:2016eto}.)

The right panel of Fig.~\ref{fig:localized}
shows the free energy excess
$\delta F \equiv F - E_{\rm vac}$
of these various solutions as a function of temperature.
Different curves have the same meaning as in the left panel.
The upper and lower red dotted curves show the free energy excess
of small and large Schwarzschild-AdS black holes, respectively.
The black dot at $\delta F \eq 0$ indicates the phase transition point.
The inset plot magnifies the region around the $\ell = 1$
bifurcation.
One sees from this figure that the
localized black hole free energy is a convex function of temperature
(as is the small Schwarzschild-AdS black hole branch).
Since $\CV = -T \, \partial^2 F/\partial T^2$,
this implies that the localized $S^8$ horizon topology black holes
have negative heat capacity,
just like undeformed small Schwarzschild-AdS black holes.

Dias, Santos, and Way interpret the energy $E_R$ as a transition point,
within the microcanonical ensemble, to a phase of spontaneously broken
$R$-symmetry with properties characterized by their localized black hole
solutions.
Dynamically, they suggest that Schwarzschild-AdS$_5 \times S^5$ black holes
which reach the $\ell \eq 1$ instability threshold (\ref{eq:E*})
should subsequently evolve toward these static localized $S^8$ horizon
topology solutions in the manner described above.

\section 	{Evolution scenarios}
\label		{sec:scenarios}

The above scenario, with evolution asymptoting to the
localized black hole solutions of Dias et al.~\cite{Dias:2016eto}
at a temperature $T_*^{\rm loc}$ greater than $\Tc$,
is consistent with currently known information.
But there are alternative possibilities which also
warrant consideration.
There may be other possible final states,
not currently known, with yet higher entropy.
Or, the presumption of asymptotic stationarity may be false,
with dynamical evolution never settling down to a well-defined
equilibrium state.
In this section, we examine various open questions and issues
associated with each of these three logical possibilities.

\subsection	{Known localized black hole as final state?}

One reason to expect the scenario of 
evolution toward the known localized black hole solutions
to be correct is the obvious: the successful construction
\cite{Dias:2016eto} of these solutions for energies
up to and a little beyond $E_*$, with the demonstration
that these localized black holes have higher entropy
than the corresponding Schwarzschild-AdS black holes.
There are, however, multiple curious or puzzling features
implied by this scenario.
One may wonder how this evolution can be
consistent with conservation of $R$-charge.
A final temperature $\Tf > \Tc$ is at odds with the
conventional picture of re-thermalization after spinodal
decomposition discussed in section \ref{sec:1stord}.
Why the difference, and should this have been expected?
This scenario implies a microcanonical description of
equilibrium states which is fundamentally different from the
canonical description of equilibrium states, despite
the fact that the large $\Nc$ limit can be viewed
as a thermodynamic limit \cite{Haan:1981ks}.
Again, why?
Finally, for finite values of $\Nc$,
what does this scenario predict for the
fate of supercooled plasma with energy
between $E_*$ and $E_{\rm c}^+$
(in the absence of further cooling)?

We begin with the disconnect between the 
description of $\Nfour$ SYM equilibrium states in the
canonical and microcanonical ensembles.
In the putative microcanonical picture,
if the total energy $E$ is less than $E_R$
(and greater than $E_{\rm vac}$ by an $O(\Nc^2)$ amount),
then in equilibrium the $SO(6)$ $R$-symmetry is spontaneously
broken down to $SO(5)$
and the heat capacity $\CV$ is negative.
In contrast,
in the canonical description
when the total energy $E$ is less than $E_{\rm c}^+$
(and greater than $E_{\rm vac}$),
then the equilibrium state is a statistical mixture 
of the confined and deconfined pure phases at $\Tc$,
and the heat capacity
$
    \CV
    = \partial E/\partial T
    = \langle (E-\langle E\rangle)^2 \rangle / T^2
$
is necessarily positive.
Nowhere in the canonical description of equilibrium states
is there any sign of spontaneous breaking of $R$-symmetry.

Substantial differences between microcanonical and canonical
descriptions are to be expected in small systems where fluctuations
in thermodynamic quantities can be substantial and the presence
or absence of energy transfer to some external environment may
play a significant role.
But in the usual thermodynamic limit,
the infinite system can itself serve as a heat bath
for any subsystem of interest.
Equivalence of ensembles is a basic result 
directly linked to the vanishing variance (\ref{eq:cluster2})
of spatially averaged observables.

We are considering a finite volume system,
$\Nfour$ SYM on a spatial three-sphere,
in the large $\Nc$ limit.
As $\Nc \to \infty$,
the number of degrees of freedom diverges
and thermodynamic functions can develop non-analyticities
(i.e., phase transitions).
Large $\Nc$ factorization \cite{Witten:1979pi,Coleman:1980nk,Yaffe:1981vf},
\begin{equation}
    \lim_{\Nc \to \infty}
    \left[
    \left\langle
	\mathcal O^2
    \right\rangle
    -
    \left\langle
	\mathcal O
    \right\rangle^2
    \right]
    = 0 \,,
\label{eq:factorize}
\end{equation}
may be viewed as the direct analog of the integrated form
(\ref{eq:cluster2}) of cluster decomposition.
Here, $\mathcal O$ is any ``classical'' large $\Nc$ observable
\cite{Yaffe:1981vf}, such as a fixed product of single trace
operators normalized to have $\O(\Nc^0)$ expectation values.
Just as with cluster decomposition, large $N$ factorization
does not hold automatically for all states in
the large $\Nc$ limit,
but rather is a diagnostic for whether a given state,
as probed by ``classical'' operators,
is indistinguishable from a statistical mixture of
states which do satisfy large $\Nc$ factorization
\cite{Yaffe:1981vf,Brown:1985yn}.
In theories with a holographic dual,
only those states satisfying large $\Nc$ factorization
will have a dual description involving a single
classical geometry.
States defined by extremization
(e.g., minimizing the free energy $F[\rho]$ or 
maximizing entropy $S[\rho]$)
will satisfy large $\Nc$ factorization 
whenever the extremum is unique.

The mismatch between canonical and microcanonical descriptions
of $\Nfour$ SYM equilibrium states 
arises from the apparent absence
of any analog of phase-separated equilibrium states,
coexisting with the pure phases at $\Tc$.
In our large $\Nc$ limit, such (hypothetical)
``phase-separated'' equilibrium states would be states
with energy intermediate between
$E_{\rm c}^+$ and $E_{\rm c}^-$,
negligible energy fluctuations,
a horizon temperature equal to $\Tc$,
and (in a Euclidean description)
expectation values of Polyakov loops, possibly multiply wound,
indicative of an eigenvalue distribution for the gauge field
holonomy in the time direction which is neither gapped nor
$\mathbf Z_{\Nc}$ invariant \cite{Aharony:2003sx}.
If such a state is a co-existing equilibrium state at $\Tc$,
then (the $O(\Nc^2)$ part of) its free energy must equal
$E_{\rm vac}$, implying that the entropy of such a state
is directly related to its energy,
$
    S = (E{-}E_{\rm vac})/\Tc
    = \frac{2\pi L}{3} \, (E{-}E_{\rm vac})
$.
This entropy would be a substantially \emph{larger} than
that of the localized black holes shown in Fig.~\ref{fig:localized}.

In the usual large volume thermodynamic limit,
it is spatial locality
(i.e., sufficiently rapid decrease of interaction strength with
distance)
which guarantees the existence of phase-separated equilibrium states
in both canonical and microcanonical descriptions.
A domain wall separating regions resembling pure phases will have a
free energy excess proportional to its area, but this excess makes
a vanishing contribution to the volume average of the free energy
density (or internal energy density) in the infinite volume limit.
Sufficiently far from the domain wall, on either side,
expectation values of local observables will be indistinguishable
from those in the corresponding pure phase.

In large $\Nc$ matrix models (or gauge theories),
correlations and fluctuations
in the space of eigenvalues should be regarded as the analog of spatial
correlations and fluctuations in a typical statistical theory.
But there is no direct analog of spatial locality as seen,
for example, in the logarithm of the Vandermonde determinant which
appears in every matrix model --- every eigenvalue interacts
with every other eigenvalue.
So features of large volume
thermodynamic limits which inextricably rely on spatial locality
may simply have no analog in large $\Nc$ thermodynamic limits.

A negative heat capacity, in a microcanonical ensemble,
is closely related to the absence of phase-separated equilibrium states.
Recall that first and second derivatives of the entropy $S(E)$ determine
the temperature $T$ and heat capacity $\CV$ via
\begin{equation}
    \frac{\partial S}{\partial E}
    =
    \frac 1T \,,
    \qquad
    \frac{\partial^2 S}{\partial E^2}
    =
    - \frac 1{T^2} \, \frac {\partial T}{\partial E}
    =
    - \frac {\CV^{-1}}{T^2} \,.
\end{equation}
A negative heat capacity 
implies that the entropy violates the usual concavity relation
(i.e., $\partial^2 S/\partial E^2 < 0$), and is instead convex.
If the microcanonical entropy $S(E)$ is convex for some range of energies,
say $E_1 < E < E_2$,
then $S(E)$ lies below its concave hull --- 
which in this interval is the straight line 
$S_{\textrm {concave-hull}}(E) = (1{-}x) \, S(E_1) + x \, S(E_2)$
with $x \equiv (E-E_1)/(E_2-E_1)$.
So convexity of the microcanonical entropy in this interval amounts to
an assertion that it is \emph{impossible} to construct any state of the
system in which some fraction $x$ of the degrees of freedom behave like a
low energy equilibrium state with entropy/energy ratio $S(E_1)/E_1$
while the complementary fraction $1{-}x$ of the degrees of freedom
behave like a high energy equilibrium state with entropy/energy
ratio $S(E_2)/E_2$,
with negligible interaction between the two subsets. 
Such a state, if it exists, would have an entropy which lies on the
above straight line connecting $S(E_1)$ and $S(E_2)$.
In ordinary large volume thermodynamic limits, 
phase-separated equilibrium states illustrate exactly
this sort of partitioning of degrees of freedom.
They play an essential role in ensuring concavity of the entropy.

In large $\Nc$ matrix models, the non-locality of interactions
in ``eigenvalue space'' may preclude any analogous partitioning
of degrees of freedom into nearly independent subsets.
Nevertheless, it is the case that eigenvalues which are farther
apart interact more weakly than eigenvalues which are closer together.
Asplund and Berenstein \cite{Asplund:2008xd}
and more recently Hanada and Maltz \cite{Hanada:2016pwv}
have argued that one should view a small AdS black hole,
localized on the $S^5$,
as representing states of $\Nfour$ SYM in which
a subset of the eigenvalues of the scalar fields are clumped together,
while the complementary set of eigenvalues are widely dispersed.
These papers are largely qualitative,
but Ref.~\cite{Hanada:2016pwv} successfully reproduces the
scaling properties of localized $S^8$ horizon topology BHs
in the limit of very small energy.
These arguments rely on an
approximate notion of locality in the $\mathbb R^6$
eigenvalue space (of the six SYM scalar fields)
replacing ordinary spatial locality.
Except for the substitution of eigenvalue space
for ordinary space,
the overall picture is highly reminiscent
of the construction of phase-separated states.

This suggests that it might, to some degree,
make sense to view
small black holes localized on the $S^5$ as 
analogs of phase-separated states associated with a first
order transition.
However, both the temperature and the
free energy of these localized black hole states are higher
than those of the pure phases at $\Tc$.
This difference could be viewed as indicating that
the approximate degree of eigenvalue locality is insufficient
to make the free energy excess associated with an ``interface''
in eigenvalue space subdominant, in the large $\Nc$ limit,
compared to the pure phase free energies.
Moreover, it is not clear, at least to this author, whether this
suggestion that localized BHs are analogs of phase-separated states
(albeit with ``extensive'' interface free energy)
is compatible with the above-mentioned characteristic of putative
phase-separated states:
that Polyakov loop expectation values interpolate
between confined and deconfined phase expectation values
in a manner indicative of a
$\mathbb Z_{\Nc}$ non-invariant, non-gapped
eigenvalue distribution for the temporal holonomy.

Turning to other issues,
it should be noted that
the initial state of supercooled plasma at the metastability
limit is invariant under the $SO(6)_R$ symmetry.
So, in the absence of any $R$-symmetry violating perturbations,
the final state must also be symmetric under $SO(6)_R$ and hence
cannot literally be described by a single geometry with a black hole
localized on the $S^5$.
This, however, is not a real problem; it just means that the
putative final state is a statistical mixture of localized
black hole states averaged over all positions on the five-sphere.
It may seem strange that a state described by a single
classical geometry (in the large $\Nc$ limit) could evolve
into a state whose dual description involves a statistical mixture
of geometries.
In particular, 
this means that an initial state satisfying large $\Nc$
factorization (\ref{eq:factorize}) evolves into a final state
which violates large $\Nc$ factorization.
However, this is a typical feature of quantum dynamics whenever
a system undergoes spontaneous symmetry breaking;
negligible quantum fluctuations can become amplified and produce
a superposition of classically distinct states which are
indistinguishable from a statistical mixture.

A more significant and puzzling feature of this
scenario concerns the fate, at large but finite $\Nc$,
of supercooled plasma with energy above the metastability limit
at $E_*$.
As discussed above, 
Schwarzschild-AdS black holes with energy between $E_*$
and $E_{\rm c}^+$ provide the dual description
of such supercooled states.
These states are locally stable (for $E > E_*$) but
one would expect non-perturbative tunneling or nucleation
instabilities to exist with decay rates which vanish
exponentially as $\Nc \to \infty$.
(After all, such an instability, however slow,
is what distinguishes the supercooled state
from true equilibrium plasma at $T \ge \Tc$.)
Viewed as a dynamical process at fixed energy,
what is the endpoint of non-perturbative decay of
supercooled plasma with energy $E > E_*$?

In a normal large volume thermodynamic limit,
a locally stable supercooled phase can decay via nucleation
and growth of critical size bubbles \cite{Langer:1969bc}.
Once one or more critical bubbles of the low temperature phase
are formed,
they expand via conversion of latent heat into 
kinetic energy of expanding bubble walls,
with subsequent bubble wall collisions leading to re-thermalization.
Bubble nucleation rates depend on the amount of supercooling
but are independent of the overall spatial volume
(in the large volume limit),
again reflecting the spatial locality of interactions.
The endpoint of bubble nucleation events in the
supercooled phase will be a phase-separated
state with the same energy but greater entropy
than the initial supercooled state
(assuming the initial energy $E > E_{\rm c}^-$).

In contrast, in large $\Nc$ SYM, the rate for a single Polyakov loop
eigenvalue to pass through a free energy barrier via
tunneling or thermal activation is $O(\Nc)$, since every
eigenvalue interacts with every other one.%
\footnote
    {%
    The $O(\Nc)$ scaling of the free energy barrier
    for a single eigenvalue may be seen explicitly
    in the weak-coupling results of
    Refs.~\cite{Aharony:2003sx,Aharony:2005bq}.
    }
Hence, unlike the situation in large volume thermodynamic limits,
the decay rate of a locally stable supercooled phase
(in finite spatial volume, at large $\Nc$)
should vanish exponentially as $\Nc \to \infty$.
Nevertheless,
it is interesting to consider dynamics of the system
at large but finite values of $\Nc$.
In the gravity-side description, a Schwarzschild-AdS black hole will
Hawking radiate, on an $O(\Nc^2)$ time scale,
until it is in equilibrium with (super)gravitational
radiation at the horizon temperature.
Since the black hole heat capacity is $O(\Nc^2)$, while
that of the radiation is $O(1)$, the black hole retains
all but an $O(\Nc^{-2})$ fraction of the total energy.
In the dual QFT,
this process looks like formation of dilute gas of glueballs
mixed in (and weakly interacting) with the deconfined plasma,
rather analogous to Coulomb plasmas in which there is no sharp
distinction between ``bound'' atoms and ``free'' ions.
Given that all radiation is trapped by the asymptotic AdS boundary
conditions, the net effect of including Hawking radiation is
merely to produce relative $O(\Nc^{-2})$ corrections to
thermodynamic quantities.
But what happens to this state,
whose gravitational description is a small black hole
in equilibrium with its radiation,
on exponentially long time scales?
Given the existence of a free energy barrier separating
this state from confined phase states with much lower
free energy, one expects the presence
of some non-perturbative instability,
but what could be the endpoint of such an instability?

As shown in Fig.~\ref{fig:localized},
the entropy of the localized BHs of Ref.~\cite{Dias:2016eto}
appears to cross that of
Schwarzschild-AdS BHs at an energy $E_R$ (\ref{eq:E_R})
which is well below $E_{\rm c}^+$.
This is based on an extrapolation of available numerical data but,
examining the quality of the fit and its extrapolation
in Fig.~\ref{fig:localized}, 
it is very hard to believe that this branch of solutions
has an entropy which remains above that of the Schwarzschild-AdS BH
all the way up to $E_{\rm c}^+$.
Assuming that the extrapolation shown in Fig.~\ref{fig:localized}
is qualitatively correct,
this scenario lacks any description of a possible endpoint for 
non-perturbative decays of supercooled plasma with energy
in the range $E_R \le E < E_{\rm c}^+$.

\subsection	{Alternative final states?}

Might there be other final states, not currently known,
with higher entropy than the known localized BH solutions?
This is a conceivable possibility,
but at this point it is pure speculation.
However, it is worth noting that stability of the localized
black holes of Ref.~\cite{Dias:2016eto}
has not been fully explored.
In particular, perturbations involving IIB supergravity fields
other than the metric and the self-dual five-form are
compatible with the symmetries of the localized BH solutions
but have not yet been studied.
So, for example, localized ``fuzzy'' black holes with non-zero
values of the three-form fields may well exist.
It is conceivable that such solutions could have horizon
temperatures precisely equal to $\Tc$,
clearly implying an interpretation as phase-separated
equilibrium states,
although it is hard to see what would pick out this particular value.

Perhaps the strongest reason for suspecting that higher entropy
solutions remain to be identified is the last puzzle mentioned above:
the lack of any identified endpoint toward which non-perturbative
decays of Schwarzschild-AdS black holes with energy
between $E_R$ and $E_{\rm c}^+$ could asymptote.

\subsection	{No stationary final state?}

A final scenario to consider is the possibility that
supercooled $\Nfour$ SYM plasma, upon reaching the limit of
metastability and becoming locally unstable,
fails to re-equilibrate.%
\footnote
    {%
    More precisely, fails to re-equilibrate on a time scale
    which remains bounded as $\Nc \to \infty$.
    }
In other words, the possibility that a Schwarzschild-AdS
black hole with energy $E < E_*$, when infinitesimally perturbed,
becomes a time dependent solution which fails to ever settle down.
Although unexpected in a strongly coupled relativistic field theory,
such behavior would be reminiscent of many body localization,
a phenomena of current interest in condensed matter physics
in which systems of interacting particles develop unusual
correlations which prevent thermalization 
\cite{PhysRevB.82.174411,
PhysRevB.83.094431,
doi:10.1146/annurev-conmatphys-031214-014726,
Chandran:2013xwa}.

In this hypothetical scenario, 
the evolution of an unstable small BH, slightly perturbed,
surely begins as described earlier in section \ref{sec:smallBH}:
the unstable mode grows and the
horizon becomes increasingly distorted on the $S^5$.
It is quite plausible that a cascade of subsequent instabilities
will lead to the development of horizon structure in a manner
similar to  Gregory-Laflamme dynamics in asymptotically flat
space \cite{Lehner:2010pn,Lehner:2011wc}.
But does this process reach the string scale 
in finite time (as seen from the boundary)?
Does the subsequent string theory dynamics then \emph{exit}
the stringy domain with different horizon topology
on an $O(1)$ time scale?
At the moment,
these portions of the ``widely expected'' scenario are conjectural.
It is certainly conceivable that these expectations are incorrect.

A critical difference between asymptotically flat and asymptotically
AdS gravitational dynamics is the impossibility,
in the asymptotic AdS case,
for a time-dependent black hole to lose energy (or angular momentum)
to radiation which propagates ever-outward with negligible subsequent
effect on the horizon dynamics.
As is well known (and fundamental to AdS/CFT duality),
the asymptotic AdS boundary acts like a box;
outward propagating gravitational waves reflect off the boundary
and return to the interior of the spacetime.
How this affects cascading horizon instabilities is far from clear.
Perhaps the horizon develops chaotic or fractal structure which
never reaches the string scale (in $O(1)$ time).%
\footnote
    {%
    Although the recent paper \cite{Bosch:2017ccw}
    examining similar dynamical issues in an asymptotically AdS
    model of exotic hairy black branes
    does support the expectation that string-scale structure
    will appear in finite time.
    }
Or perhaps the geometry does reach the string scale, but the
subsequent dynamics remains both stringy and time-dependent 
(for all $O(1)$ times),
and fails to return to the classical GR domain.

Based on experience with other examples of horizon dynamics
it would be very surprising if this scenario turned out to be correct,
as gravitational dynamics, when supplemented with a boundary
condition of regularity on a future event horizon, is effectively
dissipative.
But this intuition is based on studies of gravitational dynamics
with smooth geometries which are always far from the string scale.
Moreover, the specific problem under consideration is inherently
highly non-generic since the initial state is very specific:
supercooled $\Nfour$ SYM plasma.
Hence, the possibility of some non-generic long time
behavior is at least worth considering.

From both analytic and numerical studies of perturbations
of AdS spacetime,
there is strong evidence that there are regions of initial
conditions with non-zero measure for which the evolution
does not lead to horizon formation,
and complementary regions, also of non-zero measure,
for which horizons do form.
(See the review \cite{Martinon:2017ppj} and references therein.)
Much of this work considers initial conditions for which the
total energy $E = E_{\rm vac} \, (1 + \epsilon)$,
with $\epsilon$ infinitesimal.
In other words, the excitation energy $E {-} E_{\rm vac}$
is $O(\Nc^2)$ but tiny compared to the vacuum energy.
Most research in this area has been restricted
to spherical symmetry and work to date has focused
on asymptotically AdS spacetimes without
additional compact dimensions.
Consequently,
how this partitioning of phase space into stable and unstable
regions deforms as the excitation energy increases,
particularly in the presence of additional compact dimensions,
is not well understood.
Nevertheless, it is quite plausible that
in the asymptotically AdS$_5 \times S^5$ case
both collapsing and non-collapsing regions continue to be
present (with non-zero measure) for energies $E < E_{\rm c}^+$.
Assuming so, consider approaching the separatrix between
collapsing and non-collapsing regions of initial data
from the collapsing side.
Continuity, at finite times, with respect to small changes
in initial conditions implies that the horizon ring-down time
(i.e., the inverse of the lowest quasinormal mode frequency)
must diverge in this limit.
Hence, solutions with initial data approaching this separatrix
would be examples of dynamical black holes
which fail to equilibrate on any $O(1)$ time scale.%
\footnote
    {%
    Given that solutions in the non-collapsing region have
    no horizon, one might expect finite time continuity with
    respect to changes in initial data to also imply
    vanishing of the horizon area (and hence entropy)
    of solutions in the collapsing region as the separatrix
    is approached.
    But this is fallacious, as it ignores
    the teleological nature of event horizons.
    }
It is conceivable that infinitesimal perturbations of
unstable Schwarzschild-AdS black holes correspond to
just such initial data.

However improbable this scenario appears,
it offers a possible answer to the apparent non-existence
of any stationary solution which could represent the endpoint
of non-perturbative decays of
supercooled but locally stable plasma
(i.e., Schwarzschild-AdS black holes
with $E_* < E < E_{\rm c}^+$).

\section	{Concluding remarks}
\label		{sec:conclusion}

At the very least, the above discussion of multiple possible
scenarios for the dynamical evolution of unstable small
Schwarzschild-AdS black holes 
should make clear that interesting open questions remain
concerning the dynamics of supercooled SYM plasma beyond
the limit of metastability.
It is surely worthwhile to investigate the possible
existence of further stationary supergravity solutions,
particularly solutions which go beyond the metric plus
self-dual five-form ansatz which suffices for
the lumpy and localized BH solutions of Dias
et al.~\cite{Dias:2015pda,Dias:2016eto}.
Efforts to study numerically the time evolution of 
(slightly perturbed) unstable Schwarzschild-AdS black holes
have, so far, proven frustratingly difficult \cite{BCY}.
Even with improved numerical methods it will, at best,
be possible to follow the evolution for only a limited period
of time if fractal-like horizon structure develops.

Many of the conceptual issues discussed above also arise if
one considers the dynamics of supercooled states in large $\Nc$
pure Yang-Mills theory on $\mathbb R \times S^3$.
If the radius of the three-sphere is small compared
to the inverse strong scale $\Lambda^{-1}$,
then a weak coupling analysis is possible
\cite{Sundborg:1999ue,Aharony:2003sx}.
For small but non-zero effective coupling,
a calculable free energy barrier separates the confined
and deconfined pure phases at $\Tc$
\cite{Aharony:2005bq}.
The deconfined phase continues as a local minimum in the
free energy for temperatures below $\Tc$
down to $T_* = \Tc - (T_{\rm H} - \Tc)$,
where $T_{\rm H}$ is the Hagedorn temperature
(or the limit of metastability of the superheated
low temperature phase).
There is no sign of any additional phase-separated equilibrium
states coexisting at $\Tc$,%
\footnote
    {%
    This assumes that the key result of
    Ref.~\cite{Aharony:2005bq} is correct.
    The author has no reason to doubt the validity of this work,
    but is unaware of any independent confirmation.
    }
and there is no internal global symmetry analogous to the $SU(6)_R$ symmetry
of SYM which could spontaneously break.%
\footnote
    {%
    Moreover,
    for both pure YM on a small ($L \ll \Lambda^{-1}$)
    spatial three-sphere and conformal $\Nfour$ SYM on any size
    $S^3$, spontaneous
    breaking of the spatial $SO(4)$ symmetry seems exceedingly implausible
    as $L$ is the only relevant spatial scale.
    There is no evident shorter spatial scale which
    might play a distinct role as a characteristic de Broglie wavelength
    of excitations or a domain wall width.
    }
Hence, there is no known equilibrium state with the
same energy as the supercooled plasma at the limit of
metastability, to which the unstable supercooled system
might subsequently equilibrate.

It would be very surprising if the qualitative fate of
supercooled large $\Nc$ non-Abelian plasma,
upon reaching the limit of metastability,
is profoundly different depending on whether the plasma
is pure glue or maximally supersymmetric.
Largely because of this, if the author were a betting person,
he would wager on the last scenario discussed above:
failure to re-thermalize with no stationary final state.

\acknowledgments
    {%
    This work was supported, in part,
    by the U.~S.~Department of Energy grant DE-SC\-0011637.
    Discussions with Alex Buchel and Paul Chesler
    were responsible for the instigation of this work;
    the sharing of their insights and perspectives is
    gratefully acknowledged, as are helpful conversations
    or correspondence with Andreas Karch, Oscar Dias,
    Jorge Santos, David Berenstein, and Masanori Hanada.
    The hospitality of the University of Regensburg 
    and generous support from the Alexander von Humboldt foundation 
    during the completion of this work are also gratefully acknowledged.
    }

\bibliographystyle{JHEP}
\bibliography{globalbh}

\providecommand{\href}[2]{#2}\begingroup\raggedright\begin{thebibliography}{10}

\bibitem{Maldacena:1997re}
J.~M. Maldacena, {\it The large {N} limit of superconformal field theories and
  supergravity},  {\em Adv. Theor. Math. Phys.} {\bf 2} (1998) 231--252,
  [\href{http://xxx.lanl.gov/abs/hep-th/9711200}{{\tt hep-th/9711200}}].

\bibitem{Witten:1998qj}
E.~Witten, {\it Anti-de {Sitter} space and holography},  {\em Adv. Theor. Math.
  Phys.} {\bf 2} (1998) 253--291,
  [\href{http://xxx.lanl.gov/abs/hep-th/9802150}{{\tt hep-th/9802150}}].

\bibitem{Gubser:1998bc}
S.~S. Gubser, I.~R. Klebanov, and A.~M. Polyakov, {\it Gauge theory correlators
  from non-critical string theory},  {\em Phys. Lett.} {\bf B428} (1998)
  105--114, [\href{http://xxx.lanl.gov/abs/hep-th/9802109}{{\tt
  hep-th/9802109}}].

\bibitem{Witten:1998zw}
E.~Witten, {\it Anti-de {Sitter} space, thermal phase transition, and
  confinement in gauge theories},  {\em Adv. Theor. Math. Phys.} {\bf 2} (1998)
  505--532, [\href{http://xxx.lanl.gov/abs/hep-th/9803131}{{\tt
  hep-th/9803131}}].

\bibitem{Hubeny:2002xn}
V.~E. Hubeny and M.~Rangamani, {\it {Unstable horizons}},  {\em JHEP} {\bf 05}
  (2002) 027, [\href{http://xxx.lanl.gov/abs/hep-th/0202189}{{\tt
  hep-th/0202189}}].

\bibitem{Buchel:2015gxa}
A.~Buchel and L.~Lehner, {\it {Small black holes in $AdS_5\times S^5$}},  {\em
  Class. Quant. Grav.} {\bf 32} (2015), no.~14 145003,
  [\href{http://xxx.lanl.gov/abs/1502.1574}{{\tt arXiv:1502.1574}}].

\bibitem{Dias:2015pda}
O.~J.~C. Dias, J.~E. Santos, and B.~Way, {\it {Lumpy AdS$_{5} \times S^{5}$
  black holes and black belts}},  {\em JHEP} {\bf 04} (2015) 060,
  [\href{http://xxx.lanl.gov/abs/1501.6574}{{\tt arXiv:1501.6574}}].

\bibitem{Banks:1998dd}
T.~Banks, M.~R. Douglas, G.~T. Horowitz, and E.~J. Martinec, {\it {AdS dynamics
  from conformal field theory}},
  \href{http://xxx.lanl.gov/abs/hep-th/9808016}{{\tt hep-th/9808016}}.

\bibitem{Peet:1998cr}
A.~W. Peet and S.~F. Ross, {\it {Microcanonical phases of string theory on
  AdS$_m \times S^n$}},  {\em JHEP} {\bf 12} (1998) 020,
  [\href{http://xxx.lanl.gov/abs/hep-th/9810200}{{\tt hep-th/9810200}}].

\bibitem{Dias:2016eto}
O.~J. Dias, J.~E. Santos, and B.~Way, {\it {Localised ${AdS_5\times S^5}$ black
  holes}},  {\em Phys. Rev. Lett.} {\bf 117} (2016), no.~15 151101,
  [\href{http://xxx.lanl.gov/abs/1605.4911}{{\tt arXiv:1605.4911}}].

\bibitem{PhysRevB.82.174411}
A.~Pal and D.~A. Huse, {\it Many-body localization phase transition},  {\em
  Phys. Rev. B} {\bf 82} (Nov, 2010) 174411,
  [\href{http://xxx.lanl.gov/abs/1010.1992}{{\tt arXiv:1010.1992}}].

\bibitem{PhysRevB.83.094431}
E.~Canovi, D.~Rossini, R.~Fazio, G.~E. Santoro, and A.~Silva, {\it Quantum
  quenches, thermalization, and many-body localization},  {\em Phys. Rev. B}
  {\bf 83} (Mar, 2011) 094431, [\href{http://xxx.lanl.gov/abs/1006.1634}{{\tt
  arXiv:1006.1634}}].

\bibitem{doi:10.1146/annurev-conmatphys-031214-014726}
R.~Nandkishore and D.~A. Huse, {\it Many-body localization and thermalization
  in quantum statistical mechanics},  {\em Annual Review of Condensed Matter
  Physics} {\bf 6} (2015), no.~1 15--38,
  [\href{http://xxx.lanl.gov/abs/1404.0686}{{\tt arXiv:1404.0686}}].

\bibitem{Chandran:2013xwa}
A.~Chandran, V.~Khemani, C.~R. Laumann, and S.~L. Sondhi, {\it {Many-body
  localization and symmetry protected topological order}},  {\em Phys. Rev.}
  {\bf B89} (2014), no.~14 144201,
  [\href{http://xxx.lanl.gov/abs/1310.1096}{{\tt arXiv:1310.1096}}].

\bibitem{Assel:2015nca}
B.~Assel, D.~Cassani, L.~Di~Pietro, Z.~Komargodski, J.~Lorenzen, and
  D.~Martelli, {\it {The Casimir energy in curved space and its supersymmetric
  counterpart}},  {\em JHEP} {\bf 07} (2015) 043,
  [\href{http://xxx.lanl.gov/abs/1503.5537}{{\tt arXiv:1503.5537}}].

\bibitem{Birrell:1982ix}
N.~D. Birrell and P.~C.~W. Davies, {\em {Quantum Fields in Curved Space}}.
\newblock Cambridge Monographs on Mathematical Physics. Cambridge Univ. Press,
  Cambridge, UK, 1984.

\bibitem{Balasubramanian:1999re}
V.~Balasubramanian and P.~Kraus, {\it {A stress tensor for anti-de Sitter
  gravity}},  {\em Commun. Math. Phys.} {\bf 208} (1999) 413--428,
  [\href{http://xxx.lanl.gov/abs/hep-th/9902121}{{\tt hep-th/9902121}}].

\bibitem{Lorenzen:2014pna}
J.~Lorenzen and D.~Martelli, {\it {Comments on the Casimir energy in
  supersymmetric field theories}},  {\em JHEP} {\bf 07} (2015) 001,
  [\href{http://xxx.lanl.gov/abs/1412.7463}{{\tt arXiv:1412.7463}}].

\bibitem{Hirano:2017rbq}
S.~Hirano, {\it {Matching renormalisation schemes in holography}},
  \href{http://xxx.lanl.gov/abs/1708.1037}{{\tt arXiv:1708.1037}}.

\bibitem{Sundborg:1999ue}
B.~Sundborg, {\it {The Hagedorn transition, deconfinement and $\mathcal N=4$
  SYM theory}},  {\em Nucl. Phys.} {\bf B573} (2000) 349--363,
  [\href{http://xxx.lanl.gov/abs/hep-th/9908001}{{\tt hep-th/9908001}}].

\bibitem{Aharony:2003sx}
O.~Aharony, J.~Marsano, S.~Minwalla, K.~Papadodimas, and M.~Van~Raamsdonk, {\it
  {The Hagedorn-deconfinement phase transition in weakly coupled large $N$
  gauge theories}},  {\em Adv. Theor. Math. Phys.} {\bf 8} (2004) 603--696,
  [\href{http://xxx.lanl.gov/abs/hep-th/0310285}{{\tt hep-th/0310285}}].

\bibitem{Aharony:2005bq}
O.~Aharony, J.~Marsano, S.~Minwalla, K.~Papadodimas, and M.~Van~Raamsdonk, {\it
  {A first order deconfinement transition in large $N$ Yang-Mills theory on a
  small $S^3$}},  {\em Phys. Rev.} {\bf D71} (2005) 125018,
  [\href{http://xxx.lanl.gov/abs/hep-th/0502149}{{\tt hep-th/0502149}}].

\bibitem{Mussel:2009uw}
M.~Mussel and R.~Yacoby, {\it {The 2-loop partition function of large $N$ gauge
  theories with adjoint matter on $S^3$}},  {\em JHEP} {\bf 12} (2009) 005,
  [\href{http://xxx.lanl.gov/abs/0909.0407}{{\tt arXiv:0909.0407}}].

\bibitem{Hawking:1982dh}
S.~W. Hawking and D.~N. Page, {\it {Thermodynamics of black holes in anti-de
  Sitter space}},  {\em Commun. Math. Phys.} {\bf 87} (1983) 577.

\bibitem{Langer:1969bc}
J.~S. Langer, {\it {Statistical theory of the decay of metastable states}},
  {\em Annals Phys.} {\bf 54} (1969) 258--275.

\bibitem{Bratteli:1996xq}
O.~Bratteli and D.~W. Robinson, {\em {Operator algebras and quantum statistical
  mechanics. Vol. 2: Equilibrium states. Models in quantum statistical
  mechanics}}.
\newblock Springer, Berlin, 1996.

\bibitem{PhysRevLett.38.993}
H.~van Beijeren, {\it Exactly solvable model for the roughening transition of a
  crystal surface},  {\em Phys. Rev. Lett.} {\bf 38} (May, 1977) 993--996.

\bibitem{PhysRevB.14.4978}
S.~T. Chui and J.~D. Weeks, {\it {Phase transition in the two-dimensional
  Coulomb gas, and the interfacial roughening transition}},  {\em Phys. Rev. B}
  {\bf 14} (Dec, 1976) 4978--4982.

\bibitem{2015JSP...159..958C}
L.~{Coquille}, {\it {Examples of DLR states which are not weak limits of finite
  volume Gibbs measures with deterministic boundary conditions}},  {\em Journal
  of Statistical Physics} {\bf 159} (May, 2015) 958--971,
  [\href{http://xxx.lanl.gov/abs/1411.3265}{{\tt arXiv:1411.3265}}].

\bibitem{Gregory:1993vy}
R.~Gregory and R.~Laflamme, {\it {Black strings and $p$-branes are unstable}},
  {\em Phys. Rev. Lett.} {\bf 70} (1993) 2837--2840,
  [\href{http://xxx.lanl.gov/abs/hep-th/9301052}{{\tt hep-th/9301052}}].

\bibitem{Lehner:2010pn}
L.~Lehner and F.~Pretorius, {\it {Black strings, low viscosity fluids, and
  violation of cosmic censorship}},  {\em Phys. Rev. Lett.} {\bf 105} (2010)
  101102, [\href{http://xxx.lanl.gov/abs/1006.5960}{{\tt arXiv:1006.5960}}].

\bibitem{Lehner:2011wc}
L.~Lehner and F.~Pretorius, {\it {Final state of Gregory-Laflamme
  instability}},  in {\em {Black holes in higher dimensions}} (G.~T. Horowitz,
  ed.), ch.~3, pp.~44--68.
\newblock Cambridge Univ. Press, Cambridge, UK, 2012.
\newblock \href{http://xxx.lanl.gov/abs/1106.5184}{{\tt arXiv:1106.5184}}.

\bibitem{Haan:1981ks}
O.~Haan, {\it {Large $N$ as a thermodynamic limit}},  {\em Phys. Lett.} {\bf
  106B} (1981) 207--210.

\bibitem{Witten:1979pi}
E.~Witten, {\it {The $1 / N$ expansion in atomic and particle physics}},  {\em
  NATO Sci. Ser. B} {\bf 59} (1980) 403--419.

\bibitem{Coleman:1980nk}
S.~R. Coleman, {\it {$1/N$}},  in {\em {17th International School of Subnuclear
  Physics: Pointlike Structures Inside and Outside Hadrons Erice, Italy, July
  31-August 10, 1979}}, p.~0011, 1980.

\bibitem{Yaffe:1981vf}
L.~G. Yaffe, {\it {Large $N$ limits as classical mechanics}},  {\em Rev. Mod.
  Phys.} {\bf 54} (1982) 407.

\bibitem{Brown:1985yn}
F.~R. Brown and L.~G. Yaffe, {\it {The Coherent State Variational Algorithm: A
  numerical method for solving large $N$ gauge theories}},  {\em Nucl. Phys.}
  {\bf B271} (1986) 267--332.

\bibitem{Asplund:2008xd}
C.~T. Asplund and D.~Berenstein, {\it {Small AdS black holes from SYM}},  {\em
  Phys. Lett.} {\bf B673} (2009) 264--267,
  [\href{http://xxx.lanl.gov/abs/0809.0712}{{\tt arXiv:0809.0712}}].

\bibitem{Hanada:2016pwv}
M.~Hanada and J.~Maltz, {\it {A proposal of the gauge theory description of the
  small Schwarzschild black hole in AdS$_5\times$S$^5$}},  {\em JHEP} {\bf 02}
  (2017) 012, [\href{http://xxx.lanl.gov/abs/1608.3276}{{\tt
  arXiv:1608.3276}}].

\bibitem{Bosch:2017ccw}
P.~Bosch, A.~Buchel, and L.~Lehner, {\it {Unstable horizons and singularity
  development in holography}},  {\em JHEP} {\bf 07} (2017) 135,
  [\href{http://xxx.lanl.gov/abs/1704.5454}{{\tt arXiv:1704.5454}}].

\bibitem{Martinon:2017ppj}
G.~Martinon, {\it {The instability of anti-de Sitter space-time}},
  \href{http://xxx.lanl.gov/abs/1708.0560}{{\tt arXiv:1708.0560}}.

\bibitem{BCY}
A.~Buchel, P.~Chesler, and L.~G. Yaffe unpublished.

\end{thebibliography}\endgroup

\end{document}